%% file: main.tex
\documentclass{article} 
\usepackage{iclr2027_conference,times}

\input{math_commands.tex}

\usepackage{hyperref}
\usepackage{url}

\usepackage{amsmath,amssymb,amsthm,mathtools}
\usepackage{booktabs}
\usepackage{tabularx}
\usepackage{enumitem}
\usepackage{microtype}
\usepackage{xcolor}
\usepackage{colortbl}
\definecolor{oursbase}{HTML}{56B4E9}
\colorlet{ours}{oursbase!20}
\usepackage{array}
\usepackage{multirow}
\usepackage{graphicx}
\usepackage{wrapfig}
\usepackage{needspace}
\usepackage{caption}
\usepackage{capt-of}
\usepackage{algorithm}
\usepackage{algpseudocode}
\usepackage{pifont}\newcommand{\cmark}{\ding{51}}
\usepackage[most]{tcolorbox}

\newtcolorbox{promptbox}[1]{
    enhanced, breakable,
    title={#1}, halign title=center, fonttitle=\bfseries,
    colbacktitle=black, coltitle=white,
    colback=white, colframe=black!25, boxrule=0.6pt, arc=2mm,
    left=3mm, right=3mm, top=2mm, bottom=2mm,
    fontupper=\ttfamily\small\raggedright,
    before upper={\setlength{\parindent}{0pt}},
}

\DeclareRobustCommand{\canopy}{\textsc{Canopy}}

\iclrpreprintcopy

\title{CANOPY: Adaptive-Granularity Evidence\\Compression for Multimodal RAG}
\author{Hyojeong Yun$^{1,*}$\thanks{$^{*}$Equal contribution.}, Jueun Kim$^{2,*}$, Wook-Shin Han$^{1,\dagger}$\thanks{$^{\dagger}$Corresponding author.} \\
$^{1}$GSAI, POSTECH\quad$^{2}$CSE, POSTECH \\
\texttt{\{hjyun,jekim,wshan\}@dblab.postech.ac.kr}
}
\date{}

\begin{document}

\makeatletter
\renewcommand{\thanks}[1]{\protected@xdef\@thanks{\@thanks\protect\canopy@thankstext{#1}}}
\def\canopy@thankstext#1{\def\@thefnmark{}\@footnotetext{#1}}
\makeatother
\maketitle

\input{0_abstract}
\input{1_introduction}
\input{2_related_work}
\input{3_method}
\input{4_experiment}

\input{5_conclusion}

\newpage

\bibliography{iclr2027_conference}
\bibliographystyle{iclr2027_conference}

\clearpage
\appendix
\input{99_appendix}

\end{document}

%% file: math_commands.tex
\usepackage{amsmath,amsfonts,bm}

\def\eqref#1{equation~\ref{#1}}

\def\1{\bm{1}}

\DeclareMathAlphabet{\mathsfit}{\encodingdefault}{\sfdefault}{m}{sl}
\SetMathAlphabet{\mathsfit}{bold}{\encodingdefault}{\sfdefault}{bx}{n}



%% file: 0_abstract.tex
\begin{abstract}
Multimodal RAG retrieves text, tables, images, and videos, but choosing a retrieval granularity does not determine how much context to retain within each item.
Coarse units include irrelevant content, while uniformly fine selection can remove context needed to interpret the evidence.
Existing compressors address this trade-off with modality-specific mechanisms, leaving open a shared procedure for adapting the retained extent region by region across heterogeneous items.
We introduce \canopy{} (Canonical Projection over Hierarchy), a framework for adaptive-granularity post-retrieval evidence compression.
\canopy{} represents retrieved items as hierarchies and uses a node encoder fine-tuned on gold evidence to score regions against the query.
Parent-relative refinement compares these scores to select multiple regions at different granularities without LLM calls for node-level pruning.
Because compression cannot recover evidence that was never retrieved, a critic requests targeted follow-up retrieval when it judges the accumulated evidence insufficient; newly retrieved items are compressed before being added.
Across five QA benchmarks over a 33M-item heterogeneous corpus, \canopy{} achieves higher average answer accuracy than the evaluated retrieval baselines.
Ablations indicate that additional retrieval drives the main accuracy gains on multi-hop QA.
In the unrouted Qwen3-VL-8B-Instruct setting, compression reduces reader-input evidence tokens by 14.2--27.7\% relative to the same iterative pipeline without compression, with comparable answer accuracy.
Project page: \url{https://canopy-project-page.github.io/}.
\end{abstract}

%% file: 1_introduction.tex
\section{Introduction}
\label{sec:intro}
\suppressfloats[t]

\begin{figure}[t]
    \centering
    \includegraphics[width=\linewidth]{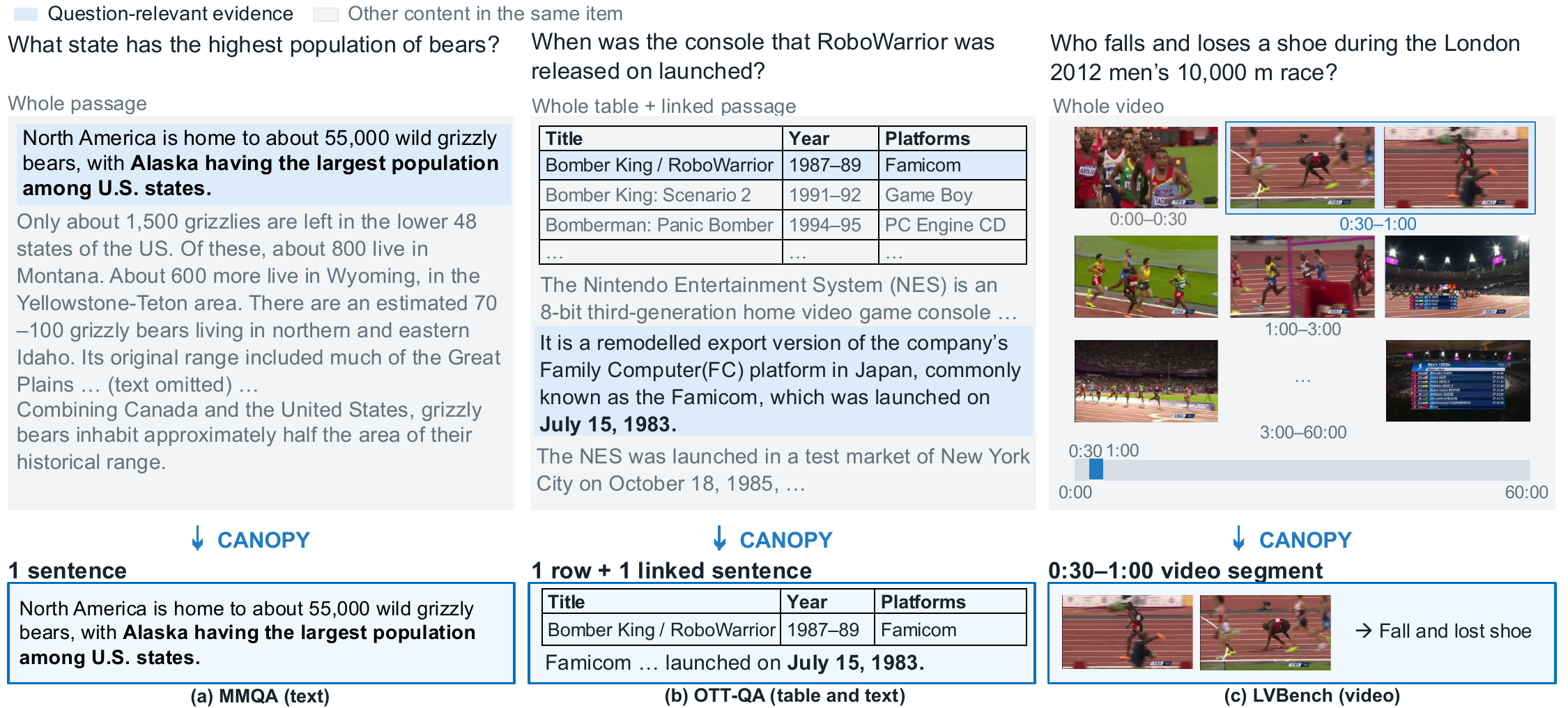}
    \caption{The evidence needed to answer a question is often only a small part of a retrieved item. \canopy{} reduces reader input by retaining a sentence (a), a table row and a linked sentence (b), or a video segment (c). Blue highlights mark question-relevant evidence; text excerpts are abbreviated and paraphrased for illustration.}
    \label{fig:motivation}
\end{figure}

Large language models remain prone to factual errors when answering questions about knowledge that is rare in, absent from, or newer than their training data~\citep{huang2025survey}.
Retrieval-augmented generation (RAG) addresses this limitation by grounding responses in external knowledge~\citep{lewis2020retrieval}, which spans text, tables, images, and videos~\citep{chen2022murag,yu2025visrag,jeong2025videorag}.
Multimodal RAG systems address how to retrieve this heterogeneous knowledge, using shared representations or modality-specific retrieval.
For example, UniversalRAG~\citep{yeo2026universalrag} routes queries to modality-specific corpora and selects among predefined retrieval granularities, such as paragraphs versus documents and clips versus full videos.

However, retrieving relevant items does not determine \emph{which parts of those items the reader needs}. Useful evidence varies in extent: a few table rows, several sentences of a passage, or a short video segment (Figure~\ref{fig:motivation}).
Passing whole items consumes limited context and can introduce distractions that impair answer quality~\citep{liu2024lost,shi2023large}.
Conversely, uniformly fine-grained selection can separate evidence from the surrounding context needed for interpretation.
Different regions within the same item may also require different amounts of context, so choosing one granularity for the item does not fully resolve this trade-off.
We therefore study \emph{post-retrieval evidence compression}: selecting concise evidence from retrieved items while adapting the retained granularity within each item.

Existing compression methods largely address this trade-off through modality-specific mechanisms.
They filter or summarize text~\citep{wang2023learning,xu2024recomp,hwang2024dslr}, retrieve relevant table schemas and cells~\citep{chen2024tablerag}, or select informative video frames~\citep{jeong2025videorag}.
This leaves open how to refine heterogeneous retrieval results through a shared mechanism.
Compression can also introduce additional inference cost when it generates summaries or invokes LLMs to select evidence.
The challenge is thus to support adaptive evidence granularity across heterogeneous items while keeping the selection procedure inexpensive.

\newsavebox{\introresultbox}
\begin{lrbox}{\introresultbox}%
\begin{minipage}{0.48\textwidth}
  \centering
  \captionsetup{type=figure,font=small,skip=4pt}
  \includegraphics[width=\linewidth]{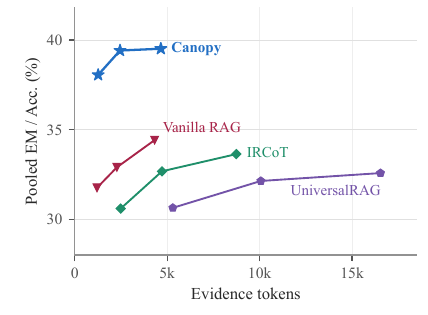}
  \caption{Pooled answer quality versus reader-input evidence tokens for complete retrieval pipelines. Points on each curve correspond to retrieval sizes $k=5,10,20$ from left to right.}
  \label{fig:topk_pooled}
\end{minipage}%
\end{lrbox}

\newlength{\introfigureintextsep}
\setlength{\introfigureintextsep}{\intextsep}
\setlength{\intextsep}{3pt}
\Needspace{\dimexpr\ht\introresultbox+\dp\introresultbox+2\intextsep+\baselineskip\relax}
\begin{wrapfigure}{r}{0.48\textwidth}
  \usebox{\introresultbox}
\end{wrapfigure}

We introduce \canopy{} (Canonical Projection over Hierarchy), a framework for adaptive-granularity evidence compression over heterogeneous retrieval results (Figure~\ref{fig:method}).
\canopy{} represents each retrieved item as a hierarchy of original regions, from the whole item to minimal fragments.
We fine-tune a node encoder with a ranking objective derived from gold evidence annotations so that query--region similarities support comparisons between parents and children.
Using these learned scores, \emph{parent-relative refinement} proceeds into every child whose query similarity matches or exceeds its parent's, discarding the remaining child subtrees.
If no child qualifies, it retains the parent. 
Because branches can stop independently, the resulting \emph{evidence forest} can retain multiple regions at different granularities within the same item.
The same refinement procedure applies to text, table, and video hierarchies without LLM calls for node-level pruning.
Images remain single nodes, and tables are retained whole for aggregate questions (Section~\ref{sec:canonical-projection}).

\setlength{\intextsep}{\introfigureintextsep}

Compression cannot, however, recover evidence that was never retrieved. For multi-hop questions, earlier evidence may reveal which information must be retrieved next~\citep{trivedi2023interleaving}.
We therefore couple hierarchical compression with critic-guided additional retrieval.
The critic examines the accumulated compressed evidence and, when it judges that evidence insufficient, identifies a missing fact and issues a targeted follow-up query.
Newly retrieved items are compressed before being added to the evidence.
The two components play complementary roles: additional retrieval seeks missing information, while compression limits the evidence volume accumulated across retrieval rounds.

We evaluate \canopy{} on NQ, HotpotQA, OTT-QA, MMQA, and LVBench over a heterogeneous corpus of approximately 33M text, table, image, and video items.
The evaluation includes initial and follow-up retrieval, rather than assuming that relevant or sufficient evidence is given.
Across the five benchmarks, \canopy{} achieves higher average answer accuracy than the evaluated retrieval baselines.
Component ablations indicate that additional retrieval drives the main accuracy gains on multi-hop QA.
In the unrouted Qwen3-VL-8B-Instruct setting, hierarchical compression reduces reader-input evidence tokens by 14.2--27.7\% relative to the same iterative pipeline without compression, with comparable answer accuracy.
Figure~\ref{fig:topk_pooled} summarizes the quality--token trade-offs of the complete pipelines across retrieval sizes.

Our contributions are threefold:
\begin{itemize}[topsep=2pt, itemsep=2pt, parsep=0pt, partopsep=0pt, leftmargin=*, labelsep=0.2em]
  \item We formulate post-retrieval evidence compression for heterogeneous multimodal RAG, focusing on adapting the retained evidence granularity within each retrieved item.

  \item We propose \canopy{}, which combines a node encoder fine-tuned on gold evidence with parent-relative hierarchical refinement to select multiple regions at different granularities, without LLM calls for node-level pruning.

  \item We evaluate \canopy{} end-to-end across five QA benchmarks over a 33M-item heterogeneous corpus, and use component ablations to distinguish the contributions of additional retrieval and compression to answer quality and reader-input evidence volume.
\end{itemize}

%% file: 2_related_work.tex
\section{Related Work}
\label{sec:related-work}
\vspace{-0.5em}
\textbf{Multimodal retrieval-augmented generation.} 
Multimodal RAG retrieves knowledge from heterogeneous sources to support answer generation.
MuRAG extends retrieval to an external memory of images and text~\citep{chen2022murag}, while VisRAG represents documents as page images to preserve layout and visual information in retrieval and generation~\citep{yu2025visrag}.
UniversalRAG additionally routes queries to modality- and granularity-specific corpora~\citep{yeo2026universalrag}.
These approaches address how sources are represented and which items are retrieved.
Selecting a retrieval granularity, however, does not by itself determine which regions within an item are useful or how much surrounding context each requires.
\canopy{} addresses this within-item decision after retrieval, complementing corpus selection and query routing rather than replacing them.

\textbf{Evidence compression and selection granularity.}
Evidence compression reduces reader input while seeking to preserve useful information.
RECOMP uses embedding-based sentence extraction or generative summarization~\citep{xu2024recomp}, while LongLLMLingua uses token probabilities from a small language model to retain question-relevant information~\citep{jiang2024longllmlingua}.
For tables, TableRAG retrieves relevant schemas and cells~\citep{chen2024tablerag}; for video, VideoRAG selects informative frames~\citep{jeong2025videorag}.
These methods already support query-dependent evidence selection, but their evidence units and selection mechanisms are tailored to their respective modalities.
Our focus is a shared procedure for adapting the retained extent region by region across heterogeneous retrieved items.

Related approaches also vary the granularity of retrieved evidence.
Mix-of-Granularity selects retrieval granularity according to the query~\citep{zhong2025mix}, while RAPTOR recursively clusters and summarizes text for retrieval at multiple abstraction levels~\citep{sarthi2024raptor}.
\canopy{} instead operates within each retrieved item, constructing a hierarchy of original regions rather than generated summaries.
Its distinction lies not only in this hierarchy but in how the node encoder is trained: gold-derived preferences between each parent and its children (Section~\ref{sec:encoder-fine-tuning}) align the learned scores with a fixed parent-relative refinement procedure, in which each parent is the local reference for which branches to explore and where to stop.
Branches stop independently, so several original regions can be retained at different granularities without a retained-unit count per item, and the same procedure applies to text, tables, and video over modality-specific fragments.
Once node embeddings are available, node-level pruning needs only similarity comparisons, without summary generation or LLM-based node selection.

\textbf{Iterative retrieval and evidence sufficiency.}
Compression can reduce retrieved content, but cannot supply evidence absent from the retrieved items.
Iterative retrieval addresses this complementary problem when intermediate findings reveal what information is needed next.
IRCoT interleaves reasoning steps with retrieval~\citep{trivedi2023interleaving}, while Self-RAG uses learned reflection tokens to assess retrieval needs, retrieved evidence, and generated responses~\citep{asai2024selfrag}.
S2G-RAG explicitly judges evidence sufficiency and remaining gaps to formulate follow-up queries, and extracts relevant sentences to manage accumulated context~\citep{li2026s2gragstructuredsufficiencygap}.
\canopy{} likewise couples evidence compression with critic-guided additional retrieval.
Its distinction lies in the compression mechanism rather than this combination itself: learned query--region scores guide parent-relative branch selection and stopping over heterogeneous items.
The same refinement procedure applies to items acquired in subsequent rounds.

%% file: 3_method.tex
\section{Method}
\label{sec:method}
\vspace{-0.5em}
\subsection{Problem Formulation}
\label{sec:problem-formulation}
\vspace{-0.5em}
Given a question $q$ and a heterogeneous corpus $\mathcal C$ containing text, tables, images, and videos, a retriever returns candidate items $X=\{x_1,\ldots,x_k\}\subseteq\mathcal C$, possibly from modality-specific corpora selected by routing.
Without within-item compression, these candidates are passed to the reader as retrieved.
In post-retrieval evidence compression, we instead construct for each item $x_i$ an evidence representation $e_i$ consisting of selected original regions, which may include the whole item, and pass $E=\{e_1,\ldots,e_k\}$ to the reader.
The objective is to reduce reader-input evidence volume while retaining information and interpretive context useful for answering $q$.

This task differs from choosing a predefined retrieval granularity.
An item's relevance does not establish which of its regions are useful or how much context each requires, and these can differ across regions of the same item, so the retained granularity must be decided within each item rather than fixed for it as a whole.
We ask: \emph{How can a shared post-retrieval procedure adapt evidence granularity within heterogeneous items to reduce reader input while preserving useful information and context?}

Evidence sufficiency is a separate concern: $X$ may lack information needed to answer $q$, which compression cannot supply; Section~\ref{sec:critic} addresses it with additional retrieval.

\begin{figure}[t]
    \centering
    \includegraphics[width=\linewidth]{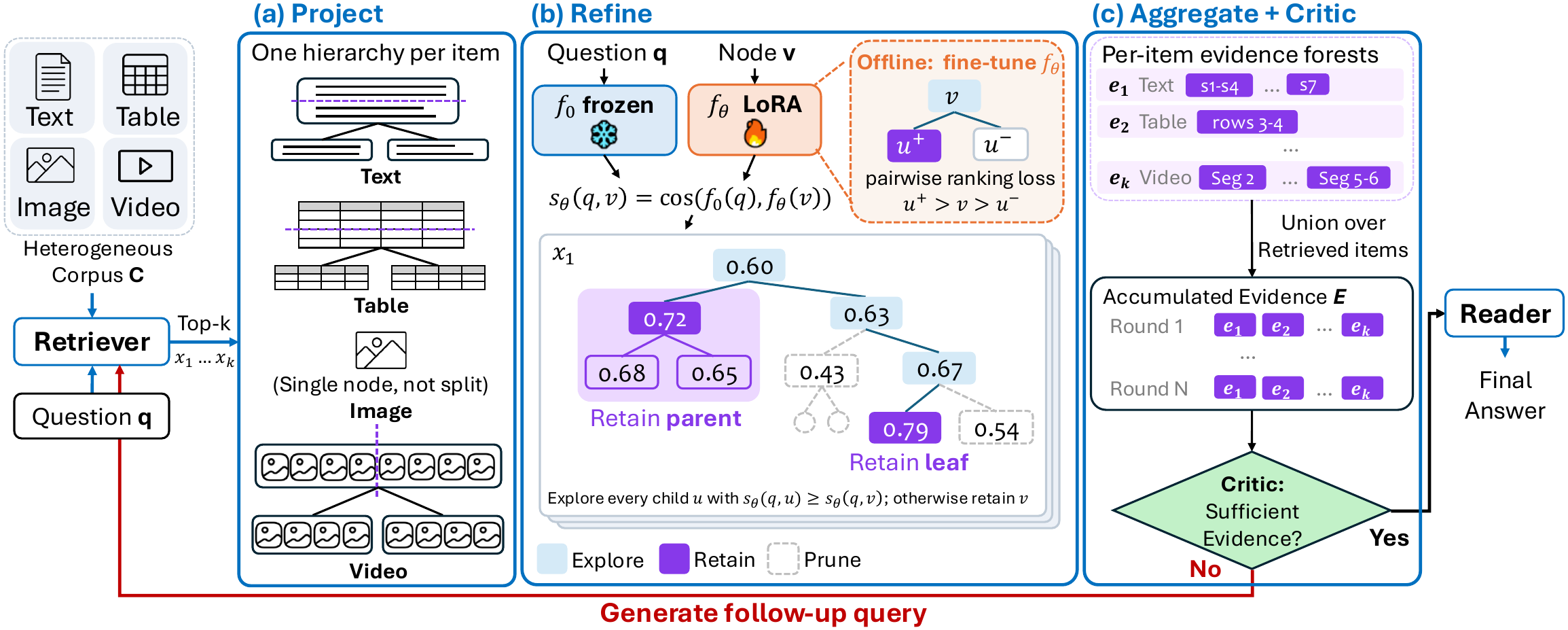}
    \caption{Overview of \canopy{}.
    (a) Each retrieved item is represented by a hierarchy of original regions, from the whole item to minimal fragments; images remain single nodes.
    (b) A frozen query encoder $f_0$ and a fine-tuned node encoder $f_\theta$ provide query--region similarity scores.
    Parent-relative refinement explores every child that scores at least as high as its parent and retains the parent if no child qualifies, allowing selected regions to lie at different depths.
    (c) The selected content is added to the accumulated evidence $E$.
    A critic assesses $E$ against the original question and guides additional retrieval when it judges the evidence insufficient.}
    \label{fig:method}
\end{figure}

\subsection{Canonical Projection over Hierarchy}
\label{sec:canonical-projection}
\vspace{-0.5em}
\canopy{} combines modality-specific hierarchy construction with a shared parent-relative refinement procedure.

\textbf{Hierarchy construction.}
Each retrieved item $x_i$ is represented by a hierarchy $\mathcal T_i$ with branching factor $B$ (Figure~\ref{fig:method}a).
Each node corresponds to an original region: the root covers the entire item, leaves correspond to minimal fragments, and intermediate nodes cover broader regions containing their descendants.
For text, each leaf is a sentence.
Tables are partitioned by row, with one row per leaf and column headers preserved at every level.
Videos are partitioned along the temporal axis into 30-second leaf segments; frames are sampled at one frame every 10 seconds, up to 32 frames.
Images remain single-node hierarchies without spatial pruning.

Aggregate questions over a table may draw on any subset of its rows, which row-level relevance cannot identify, so refinement could remove rows the answer needs.
When a table first appears among the retrieved items, an LLM therefore classifies the original question, once and from the question alone, as aggregate or lookup: aggregate questions retain the table whole, and lookup questions use hierarchical refinement.

\textbf{Learned node scores.}
Let $q_t$ denote the retrieval query at round $t$, with $q_1=q$;
subsequent queries are generated as described in
Section~\ref{sec:critic}.
For a node $v$, we compute
\begin{equation}
    s_\theta(q_t,v)
    =
    \cos\big(f_0(q_t),\,f_\theta(v)\big),
    \label{eq:node-score}
\end{equation}
where $f_0$ is the frozen pretrained query encoder and $f_\theta$ is the node encoder initialized from $f_0$.
We fine-tune $f_\theta$ using gold-evidence preferences (Section~\ref{sec:encoder-fine-tuning}) so that relative scores can guide comparisons between a region and its sub-regions.
Because the scores serve only such relative comparisons within an item, no absolute similarity threshold, which would have to be set separately for each modality, is needed.

\textbf{Parent-relative refinement.}
Refinement starts at the root of $\mathcal T_i$.
At each visited internal node $v$, we compare its score with those of its children (Figure~\ref{fig:method}b).
If at least one child $u$ satisfies $s_\theta(q_t,u)\geq s_\theta(q_t,v)$, we replace $v$ with all qualifying children, continue refinement within each of them, and discard the remaining child subtrees.
If no child qualifies, we retain $v$ without further refinement; a retained node keeps all of its fragments, including leaves that would score low on their own.
A visited leaf is retained directly.
The learned scores therefore determine which branches are explored, but do not change the traversal conditions.

Different branches can stop at different depths (Algorithm~\ref{alg:canopy} in Appendix~\ref{app:algorithm}).
The selected node set $\mathcal S_i$ contains no node that is an ancestor of another selected node, so the subtrees rooted at these nodes form an \emph{evidence forest}.
The compressed evidence $e_i$ consists of the original content covered by the selected nodes, rather than generated summaries.
An item may also remain whole, since refinement stops at the root when no child qualifies.

The retained extent follows from these comparisons alone; no retained-unit count or token budget is prescribed.
Node embeddings do not depend on the query, so they can be computed once per item and cached.
Once embeddings are available, node-level pruning requires only similarity evaluation and comparisons, without generative LLM calls.
This does not eliminate the cost of node encoding or the separate LLM calls for aggregate classification and evidence assessment.

\subsection{Critic-Guided Additional Retrieval}
\label{sec:critic}
\vspace{-0.5em}
Parent-relative refinement selects regions using relevance scores; it does not determine whether their combined content suffices to answer the question.
Moreover, required evidence may be absent from the retrieved items.
\canopy{} therefore pairs compression with an LLM critic that examines the accumulated evidence
(Figure~\ref{fig:method}c).

At round $t$, the retriever uses $q_t$ to obtain candidate items.
Newly retrieved items are compressed using $s_\theta(q_t,v)$ and added to the accumulated evidence, denoted by $E_t$; re-retrieved items are skipped.
The critic receives the original question $q$ and $E_t$, and judges whether the available evidence supports an answer.
If it judges the evidence sufficient, the reader generates an answer from $E_t$.
Otherwise, the critic identifies a missing fact and formulates a targeted follow-up query $q_{t+1}$.
Thus, compression of newly retrieved items is conditioned on the current retrieval query, while sufficiency is assessed against the original question.

The loop terminates when the critic judges the accumulated evidence sufficient or the maximum number of retrieval rounds $R$ is reached.
The critic's judgment is an operational stopping criterion rather than a guarantee of evidence completeness; reaching the round limit also does not establish sufficiency.
At round $R$, the reader answers from $E_R$ without a critic call.

\subsection{Encoder Fine-Tuning for \canopy{}}
\label{sec:encoder-fine-tuning}
\vspace{-0.5em}
Relevance scores from an off-the-shelf encoder need not support parent--child comparisons between regions of different extents.
We therefore fine-tune $f_\theta$ to express preferences derived from the locations of gold evidence (Figure~\ref{fig:frozen_vs_finetuned}).

\textbf{Notation.}
Consider a training pair consisting of a query $q$ and a gold item $x_i$ with hierarchy $\mathcal T_i$ and localized gold evidence.
For a node $v$, let $F(v)$ denote the set of minimal fragments covered by $v$ and $\mathrm{ch}(v)$ its children.
We map the gold spans described in Appendix~\ref{app:corpus} onto a gold fragment set $G_i$, containing every minimal fragment that overlaps a gold span.
A node $v$ is \emph{gold-overlapping} if $F(v)\cap G_i\neq\emptyset$ and \emph{fully gold} if $F(v)\subseteq G_i$.
Items without a located gold span are excluded from preference construction, as described in Appendix~\ref{app:corpus}.
Gold annotations are used to construct training preferences; they are not required for inference-time refinement.

\textbf{Training instances.}
Let $\mathcal V_i$ be the set of internal nodes $v\in\mathcal T_i$ that are gold-overlapping and have no fully gold proper ancestor.
Equivalently, these are the internal nodes reached by descending from the root through gold-overlapping children and stopping at fully gold nodes or leaves, that is, the nodes an ideal refinement would visit.
Each $v\in\mathcal V_i$ yields a training instance with preference pairs $\mathcal P(v)$, where $(a,b)\in\mathcal P(v)$ means that $a$ should score higher than $b$.

If $v$ is fully gold, subdivision cannot remove non-gold fragments under the mapped annotations, so we encourage retaining $v$ by preferring it over each of its children: $\mathcal P(v)=\{(v,u):u\in\mathrm{ch}(v)\}$.

Otherwise, we divide the children into gold-overlapping children
$\mathrm{ch}^+(v)=\{u\in\mathrm{ch}(v):F(u)\cap G_i\neq\emptyset\}$
and the remaining children
$\mathrm{ch}^-(v)=\mathrm{ch}(v)\setminus\mathrm{ch}^+(v)$.
We encourage refinement into the gold-overlapping children and pruning of the others through
\begin{equation}
    \begin{aligned}
        \mathcal P(v)=\;&
        \{(u,v):u\in\mathrm{ch}^+(v)\}
        \;\cup\;
        \{(v,u):u\in\mathrm{ch}^-(v)\}\\
        \;\cup\;&
        \{(u,u'):
          u\in\mathrm{ch}^+(v),\,
          u'\in\mathrm{ch}^-(v)\}.
    \end{aligned}
    \label{eq:pref-mixed}
\end{equation}

The first two sets in Equation~\ref{eq:pref-mixed} train the parent-relative comparisons used during refinement.
The third reinforces the separation between gold-overlapping and non-gold siblings.
They provide annotation-based targets for evidence selection, rather than a guarantee that all context needed for answering is annotated or preserved.

\textbf{Training objective.}
For each instance, we apply a pairwise ranking loss over
$\mathcal P(v)$:
\begin{equation}
    \ell(q,v)
    =
    \log\!\Big[
        1+
        \sum_{(a,b)\in\mathcal P(v)}
        \exp\!\big(
            \lambda[
                s_\theta(q,b)-s_\theta(q,a)
            ]
        \big)
    \Big],
    \label{eq:node-ranking-loss}
\end{equation}
where $\lambda>0$ scales score differences.
With $\mathcal D=\{(q,v):(q,x_i)\text{ is a training pair},\, v\in\mathcal V_i\}$, we minimize the mean of $\ell(q,v)$ over $\mathcal D$, updating only the LoRA~\citep{hu2021loralowrankadaptationlarge} parameters in the language backbone of $f_\theta$; the query encoder $f_0$ remains frozen.

%% file: 4_experiment.tex
\section{Experiments}
\label{sec:experiments}
\subsection{Experimental Setup}
\label{sec:setup}
\textbf{Datasets.}
We evaluate \canopy{} on five question-answering benchmarks: Natural Questions (NQ)~\citep{kwiatkowski2019natural} for single-hop text QA, HotpotQA~\citep{yang2018hotpotqa} for multi-hop reasoning across documents, OTT-QA~\citep{chen2021ottqa} for tables and text, MultimodalQA (MMQA)~\citep{talmor2021multimodalqa} for text, tables, and images, and the LVBench~\citep{wang2025lvbench} adaptation released by UniversalRAG~\citep{yeo2026universalrag} for video QA.
Each benchmark is split into a test portion and a disjoint train portion used only for node-encoder fine-tuning.
Unless noted otherwise, retrieval runs over a unified corpus of approximately 33M text, table, image, and video items built from the retrieval collections of these benchmarks; Appendix~\ref{app:corpus} gives the splits and corpus statistics.

\textbf{Baselines.}
No Retrieval uses only the reader's parametric knowledge.
Vanilla RAG retrieves from the unified corpus with a shared multimodal encoder and passes the evidence directly to the reader.
UniversalRAG~\citep{yeo2026universalrag} selects modality--granularity pairs and retrieves from the corresponding corpora.
IRCoT~\citep{trivedi2023interleaving} interleaves retrieval with chain-of-thought reasoning for at most three iterations, as in \canopy{}.
On NQ, HotpotQA, and OTT-QA, we also compare with RECOMP's extractive and abstractive compressors~\citep{xu2024recomp}, LongLLMLingua~\citep{jiang2024longllmlingua} for prompt compression, and S2G-RAG~\citep{li2026s2gragstructuredsufficiencygap} for iterative retrieval guided by evidence sufficiency and remaining gaps.
On LVBench, we compare with AKS~\citep{tang2025adaptive}, which selects query-relevant keyframes from a single video given with the question, so both methods receive the gold video.

\textbf{Metrics.}
We report exact match (EM) and token-level F1 on NQ, HotpotQA, OTT-QA, and MMQA, and accuracy (Acc.) on multiple-choice LVBench.
\#Tok is the mean number of evidence tokens per question in the reader's answer prompt, and LLM Calls is the mean number of LLM calls per question.
Lat.\ is seconds per question under batched execution, excluding vector search and node encoding (Appendix~\ref{app:implementation_details}).

\textbf{Implementation Details.}
All retrieval-based methods share Qwen3-VL-Embedding-2B~\citep{li2026qwen3vlembeddingqwen3vlrerankerunifiedframework} as the retrieval encoder; in \canopy{}, it serves as the frozen query encoder $f_0$ and initializes the node encoder $f_\theta$, which is fine-tuned with a single LoRA adapter shared across all five benchmarks (Appendix~\ref{app:implementation_details}).
Table~\ref{tab:overall_results} evaluates all methods with Qwen3-VL-8B-Instruct~\citep{bai2025qwen3} and with InternVL3.5-8B~\citep{wang2025internvl3}, and all other experiments use Qwen3-VL-8B-Instruct; within each setting, the same model handles every LLM call, including the reader, critic, router, and aggregate classifier.
Unless noted otherwise, we set $B=2$, retrieve $k=10$ items per round, and allow at most $R=3$ rounds; baseline reproduction, modality routing, context caps, and the latency measurement setup are described in Appendix~\ref{app:implementation_details}.

\subsection{Experimental Results}
\label{sec:results}
\vspace{-0.5em}

\textbf{Overall Results.}
With Qwen3-VL-8B-Instruct as the reader, \canopy{} is best in EM and F1 on the four benchmarks without video (Table~\ref{tab:overall_results}); its $+0.3$ F1 over IRCoT on HotpotQA is not significant (95\% CI $[-1.7, +2.2]$).
The gains concentrate on multi-hop HotpotQA and OTT-QA, where single-round retrieval misses evidence that becomes identifiable only after earlier evidence is read.
These gains come at a small cost in evidence volume: at 2.1 rounds on average, \canopy{} passes at most 28\% more evidence tokens than Vanilla RAG and 46--88\% fewer than UniversalRAG, since each round adds only retained regions.
Routing is complementary: it helps only on LVBench, where text queries embed far from video items in the shared space~\citep{liang2022mind,yeo2026universalrag}, so \canopy{}$^\dagger$ takes the best LVBench accuracy while unrouted \canopy{} stays ahead elsewhere.
However, with InternVL3.5-8B on LVBench, routed \canopy{} achieves lower accuracy than UniversalRAG (34.0 versus 37.2) while using more evidence tokens (31,218 versus 27,290).

\begin{table}[h]
    \vspace{-8pt}
    \captionsetup{skip=4pt}
    \centering
    \caption{Overall performance and inference efficiency over heterogeneous multimodal corpora.
    $^\dagger$Queries are routed to modality-specific corpora before retrieval, following UniversalRAG.
    Avg.\ is the mean of EM (Acc.\ for LVBench) over the five benchmarks.}
    \label{tab:overall_results}
    \footnotesize
    \setlength{\tabcolsep}{2pt}
    \renewcommand{\arraystretch}{1.15}
    \resizebox{\textwidth}{!}{%
    \begin{tabular}{@{}ll*{4}{rr>{\columncolor{gray!12}}r>{\columncolor{gray!12}}r}c>{\columncolor{gray!12}}r>{\columncolor{gray!12}}rcr@{}}
        \toprule
        \multirow{2}{*}{Model} & \multirow{2}{*}{Method}
        & \multicolumn{4}{c}{NQ} & \multicolumn{4}{c}{HotpotQA} & \multicolumn{4}{c}{OTT-QA} & \multicolumn{4}{c}{MMQA} & \multicolumn{3}{c}{LVBench}
        & \multirow{2}{*}{Avg.} & \multirow{2}{*}{LLM Calls} \\
        \cmidrule(lr){3-6} \cmidrule(lr){7-10} \cmidrule(lr){11-14} \cmidrule(lr){15-18} \cmidrule(lr){19-21}
        & & EM & F1 & \#Tok & Lat. & EM & F1 & \#Tok & Lat. & EM & F1 & \#Tok & Lat. & EM & F1 & \#Tok & Lat. & Acc. & \#Tok & Lat. & & \\
        \midrule
        \multirow{6}{*}{Qwen3-VL-8B-Instruct}
        & No Retrieval & 17.5 & 29.0 & 0 & 0.02 & 23.3 & 31.4 & 0 & 0.03 & 7.5 & 11.9 & 0 & 0.03 & 23.1 & 27.1 & 0 & 0.02 & 30.4 & 0 & 0.03 & 20.4 & 1.0 \\
        & Vanilla RAG & 38.2 & 52.6 & 1,565 & 0.08 & 38.9 & 49.7 & 1,520 & 0.07 & 12.9 & 17.5 & 2,185 & 0.11 & 40.1 & 45.9 & 1,863 & 0.09 & 34.4 & 4,236 & 0.32 & 32.9 & 1.0 \\
        & UniversalRAG & 35.0 & 50.4 & 2,804 & 0.51 & 36.5 & 46.7 & 3,980 & 0.32 & 10.5 & 14.4 & 5,546 & 0.41 & 36.9 & 42.0 & 4,131 & 0.48 & 41.8 & 33,853 & 4.23 & 32.1 & 2.0 \\
        & IRCoT & 35.0 & 51.3 & 3,179 & 0.50 & 47.4 & 61.8 & 3,363 & 0.52 & 24.3 & 30.8 & 4,707 & 0.72 & 38.9 & 46.9 & 3,817 & 0.58 & 17.8 & 8,518 & 2.08 & 32.7 & 3.6 \\
        \rowcolor{ours}\cellcolor{white} & \canopy{} & \textbf{39.7} & \textbf{54.4} & 1,512 & 0.28 & \textbf{49.6} & \textbf{62.1} & 1,816 & 0.33 & \textbf{27.7} & \textbf{33.4} & 2,796 & 0.52 & \textbf{45.1} & \textbf{51.1} & 2,123 & 0.41 & 35.0 & 3,978 & 0.93 & \textbf{39.4} & 3.0 \\
        \rowcolor{ours}\cellcolor{white} & \canopy{}$^\dagger$ & 38.1 & 52.7 & 2,715 & 0.64 & 47.6 & 60.1 & 2,967 & 0.59 & 20.1 & 24.9 & 4,929 & 0.95 & 42.1 & 47.4 & 4,251 & 1.23 & \textbf{43.4} & 26,907 & 6.54 & 38.3 & 4.9 \\
        \midrule
        \multirow{6}{*}{InternVL3.5-8B}
        & No Retrieval & 15.0 & 25.1 & 0 & 0.02 & 18.6 & 25.4 & 0 & 0.02 & 6.3 & 10.3 & 0 & 0.02 & 21.4 & 24.8 & 0 & 0.03 & 30.8 & 0 & 0.02 & 18.4 & 1.0 \\
        & Vanilla RAG & 40.4 & 53.8 & 1,581 & 0.08 & 38.6 & 48.5 & 1,527 & 0.07 & 12.6 & 17.0 & 2,185 & 0.10 & 39.8 & 45.2 & 1,910 & 0.09 & 33.6 & 7,728 & 0.62 & 33.0 & 1.0 \\
        & UniversalRAG & 33.7 & 47.8 & 4,791 & 0.46 & 38.8 & 48.6 & 4,559 & 0.26 & 10.7 & 14.3 & 5,790 & 0.33 & 33.0 & 37.5 & 6,598 & 0.49 & \textbf{37.2} & 27,290 & 4.25 & 30.7 & 2.0 \\
        & IRCoT & 35.2 & 51.3 & 2,728 & 0.32 & 44.0 & 55.9 & 2,647 & 0.30 & 16.9 & 22.3 & 3,730 & 0.41 & 37.6 & 43.5 & 3,220 & 0.37 & 10.6 & 10,541 & 1.38 & 28.9 & 2.4 \\
        \rowcolor{ours}\cellcolor{white} & \canopy{} & \textbf{41.0} & \textbf{54.4} & 1,452 & 0.27 & \textbf{48.8} & \textbf{60.1} & 1,764 & 0.33 & \textbf{21.8} & \textbf{27.0} & 2,636 & 0.50 & \textbf{42.9} & \textbf{48.6} & 2,138 & 0.42 & 34.4 & 7,111 & 1.96 & \textbf{37.8} & 3.0 \\
        \rowcolor{ours}\cellcolor{white} & \canopy{}$^\dagger$ & 39.5 & 52.1 & 2,320 & 0.54 & 47.7 & 58.5 & 2,269 & 0.48 & 13.5 & 18.0 & 3,276 & 0.69 & 38.2 & 43.3 & 5,366 & 1.27 & 34.0 & 31,218 & 8.16 & 34.6 & 4.9 \\
        \bottomrule
    \end{tabular}%
    }
    \vspace{-6pt}
\end{table}

\begin{figure}[H]
    \vspace{-2pt}
    \centering
    \includegraphics[width=0.9\linewidth]{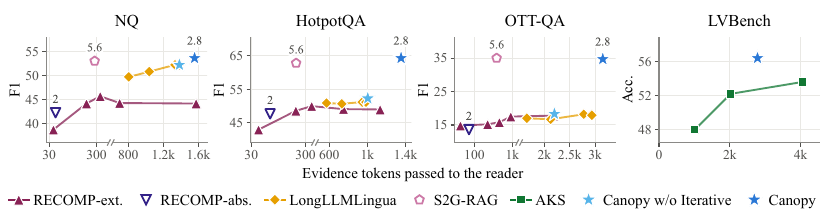}
    \caption{Answer quality versus evidence tokens for compression methods.
    Lines connect each method's budget settings in increasing order.
    Numbers give the mean LLM calls per question; hollow markers denote LLM-based baselines.}
    \label{fig:compression_budget}
    \vspace{-6pt}
\end{figure}

\textbf{Comparison with Modality-Specific Compression Methods.}
Figure~\ref{fig:compression_budget} compares \canopy{} with the compression baselines of Section~\ref{sec:setup}, retrieving over each benchmark's own collection rather than the unified corpus, and with the gold video on LVBench.
Larger budgets do not close the gap (Appendix~\ref{app:compression_budget}): LongLLMLingua saturates 1.4--16.5 F1 below \canopy{}, and AKS spends 1.5$\times$ its tokens for 2.8 fewer accuracy points.
Without iterative retrieval, \canopy{} is on par with LongLLMLingua at similar budgets while pruning with embedding similarity alone, so the wide margins on HotpotQA and OTT-QA come from iterative retrieval.
Only S2G-RAG, even with its trained judge replaced by the untrained reader LLM, reaches comparable F1 with far fewer evidence tokens, but it makes 5.6 LLM calls per question against 2.8 for \canopy{}, whose pruning within each item requires no LLM calls; \canopy{} thus trades more evidence tokens for fewer LLM calls.

\subsection{Ablation Study and Analysis}
\label{sec:ablation}
\vspace{-0.5em}

\textbf{Effects of compression and iterative retrieval.}
Table~\ref{tab:component_ablation} evaluates \canopy{} without modality routing.
Removing projection passes whole items to the reader; removing iterative retrieval restricts the pipeline to a single round.
Compression reduces evidence tokens by 14.2--27.7\% relative to \textit{w/o Projection}, with EM changes ranging from $-0.9$ to $+1.2$ points and a $+0.4$-point change in LVBench accuracy, all of whose 95\% CIs include zero ($[0.0, +2.4]$, $[-1.0, +1.8]$, $[-1.7, +1.7]$, $[-2.4, +0.7]$, and $[-2.0, +2.8]$ on NQ, HotpotQA, OTT-QA, MMQA, and LVBench).
Removing iterative retrieval lowers HotpotQA and OTT-QA EM by 10.1 and 14.9 points, respectively.
The accuracy gains thus come from additional retrieval, while compression reduces what the reader must process at comparable accuracy.
This saving is that of the whole projection, fine-tuned encoder and hierarchical selection together; Table~\ref{tab:frozen_vs_finetuned} isolates the encoder and Appendix~\ref{app:projection_strategies} the selection rule.
The routed variants are reported in Appendix~\ref{app:routed_ablation}; on routed LVBench, compression saves 33.3\% of tokens but lowers accuracy by 1.8 points.

\begin{table}[H]
    \vspace{-8pt}
    \captionsetup{skip=4pt}
    \centering
    \caption{Effect of \canopy{}'s components without modality routing.}
    \label{tab:component_ablation}
    \footnotesize
    \setlength{\tabcolsep}{2pt}
    \renewcommand{\arraystretch}{1.15}
    \resizebox{\textwidth}{!}{%
    \begin{tabular}{@{}lrr>{\columncolor{gray!12}}r>{\columncolor{gray!12}}rrr>{\columncolor{gray!12}}r>{\columncolor{gray!12}}rrr>{\columncolor{gray!12}}r>{\columncolor{gray!12}}rrr>{\columncolor{gray!12}}r>{\columncolor{gray!12}}rc>{\columncolor{gray!12}}r>{\columncolor{gray!12}}r@{}}
        \toprule
        \multirow{2}{*}{Method}
        & \multicolumn{4}{c}{NQ} & \multicolumn{4}{c}{HotpotQA} & \multicolumn{4}{c}{OTT-QA} & \multicolumn{4}{c}{MMQA} & \multicolumn{3}{c}{LVBench} \\
        \cmidrule(lr){2-5} \cmidrule(lr){6-9} \cmidrule(lr){10-13} \cmidrule(lr){14-17} \cmidrule(lr){18-20}
        & EM & F1 & \#Tok & Lat. & EM & F1 & \#Tok & Lat. & EM & F1 & \#Tok & Lat. & EM & F1 & \#Tok & Lat. & Acc. & \#Tok & Lat. \\
        \midrule
        \rowcolor{ours} \canopy{} & \textbf{39.7} & \textbf{54.4} & 1,512 & 0.28 & \textbf{49.6} & 62.1 & 1,816 & 0.33 & \textbf{27.7} & \textbf{33.4} & 2,796 & 0.52 & 45.1 & 51.1 & 2,123 & 0.41 & \textbf{35.0} & 3,978 & 0.93 \\
        w/o Projection & 38.5 & 52.9 & 1,772 & 0.30 & 49.2 & \textbf{62.2} & 2,122 & 0.35 & \textbf{27.7} & 33.3 & 3,258 & 0.55 & \textbf{46.0} & \textbf{51.8} & 2,482 & 0.42 & 34.6 & 5,503 & 1.26 \\
        w/o Iterative & 37.7 & 52.4 & 1,336 & 0.08 & 39.5 & 50.2 & 1,301 & 0.07 & 12.8 & 17.2 & 1,859 & 0.11 & 39.2 & 45.0 & 1,584 & 0.09 & 33.6 & 2,941 & 0.20 \\
        w/o Projection, Iterative (Vanilla RAG) & 38.2 & 52.6 & 1,565 & 0.08 & 38.9 & 49.7 & 1,520 & 0.07 & 12.9 & 17.5 & 2,185 & 0.11 & 40.1 & 45.9 & 1,863 & 0.09 & 34.4 & 4,236 & 0.32 \\
        \bottomrule
    \end{tabular}%
    }
    \vspace{-8pt}
\end{table}

\begin{figure}[H]
    \vspace{-6pt}
    \centering
    \includegraphics[width=0.9\linewidth]{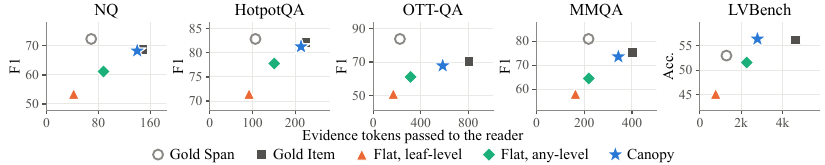}
    \caption{Answer quality (F1; accuracy on LVBench) versus evidence tokens per projection strategy, with the gold item given directly. Gold Span keeps only the annotated span.}
    \label{fig:projection_tradeoff}
    \vspace{-6pt}
\end{figure}

\textbf{Comparison of evidence selection strategies.}
To assess adaptive selection across the hierarchy, we provide gold items directly and compare \canopy{} with retaining whole items (Gold Item), annotated spans (Gold Span), and flat selection over the same hierarchy: leaf-level flat selection keeps the highest-scoring leaf, and any-level flat selection keeps the highest-scoring node, which may be the item itself (Figure~\ref{fig:projection_tradeoff}).
Both flat selectors score candidates with the retrieval encoder $f_0$, since $f_\theta$ is trained for parent-relative comparisons rather than for ranking regions against one another.
\canopy{} exceeds both flat selectors on all five benchmarks while retaining more tokens, reflecting the trade-off between evidence volume and context preservation.
Relative to Gold Item, it reduces LVBench tokens by 40.0\% with comparable accuracy (56.4 versus 56.2; 95\% CI of the difference $[-3.4, +3.8]$), whereas on OTT-QA it saves 27.5\% with an F1 decrease from 70.5 to 67.8.
Savings on NQ and HotpotQA are approximately 5\%.
Gold Span is not a performance upper bound: its LVBench accuracy is 53.0, suggesting that context beyond annotated spans can be useful.
Appendix~\ref{app:projection_strategies} repeats the comparison in the full pipeline with flat selection scored by the same $f_\theta$: its sweep over the number of retained leaves reaches quality--token points comparable to \canopy{}'s, which \canopy{} reaches through parent-relative branch selection and stopping rather than a retained-unit count chosen by sweeping.

\textbf{Sensitivity to branching factor and retrieval size.}
Table~\ref{tab:parameter_sensitivity} summarizes the sensitivity of \canopy{} without modality routing to $B$ and $k$.
A larger branching factor does not consistently improve answer quality or reduce evidence volume, while retrieving more items generally increases tokens with modest or non-monotonic quality changes.
Appendix Table~\ref{tab:branching_factor} reports the full $B$ sweep with and without routing; Appendix Table~\ref{tab:topk} reports the $k$ sweep and comparisons with retrieval baselines.

\begin{table}[H]
    \vspace{-8pt}
    \captionsetup{skip=4pt}
    \centering
    \caption{Sensitivity of \canopy{} without modality routing to branching factor $B$ and retrieval size $k$.}
    \label{tab:parameter_sensitivity}
    \scriptsize
    \setlength{\tabcolsep}{2pt}
    \renewcommand{\arraystretch}{1.1}
    \begin{tabular}{@{}ccrr>{\columncolor{gray!12}}rrr>{\columncolor{gray!12}}rrr>{\columncolor{gray!12}}rrr>{\columncolor{gray!12}}rc>{\columncolor{gray!12}}r@{}}
        \toprule
        \multirow{2}{*}{Param.} & \multirow{2}{*}{Value}
        & \multicolumn{3}{c}{NQ} & \multicolumn{3}{c}{HotpotQA} & \multicolumn{3}{c}{OTT-QA} & \multicolumn{3}{c}{MMQA} & \multicolumn{2}{c}{LVBench} \\
        \cmidrule(lr){3-5} \cmidrule(lr){6-8} \cmidrule(lr){9-11} \cmidrule(lr){12-14} \cmidrule(lr){15-16}
        & & EM & F1 & \#Tok & EM & F1 & \#Tok & EM & F1 & \#Tok & EM & F1 & \#Tok & Acc. & \#Tok \\
        \midrule
        \multirow{3}{*}{$B$} & 2 & \textbf{39.7} & \textbf{54.4} & 1,512 & 49.6 & 62.1 & 1,816 & \textbf{27.7} & \textbf{33.4} & 2,796 & 45.1 & 51.1 & 2,123 & 35.0 & 3,978 \\
         & 3 & 39.0 & 53.5 & 1,560 & \textbf{49.8} & \textbf{62.3} & 1,862 & 26.9 & 32.9 & 2,765 & \textbf{45.2} & \textbf{51.2} & 2,130 & 34.6 & 4,117 \\
         & 4 & 39.2 & 53.9 & 1,538 & 49.6 & \textbf{62.3} & 1,827 & 25.8 & 31.6 & 2,734 & 44.6 & 50.8 & 2,101 & \textbf{35.4} & 4,312 \\
        \midrule
        \multirow{3}{*}{$k$} & 5 & 37.5 & 51.9 & 756 & 49.5 & 61.2 & 946 & 26.3 & 31.6 & 1,458 & 44.0 & 49.2 & 1,106 & 33.0 & 2,051 \\
         & 10 & 39.7 & 54.4 & 1,512 & 49.6 & 62.1 & 1,816 & \textbf{27.7} & \textbf{33.4} & 2,796 & \textbf{45.1} & \textbf{51.1} & 2,123 & \textbf{35.0} & 3,978 \\
         & 20 & \textbf{40.3} & \textbf{55.8} & 2,982 & \textbf{50.7} & \textbf{63.0} & 3,555 & 27.6 & 33.2 & 5,362 & 44.0 & 50.3 & 4,064 & \textbf{35.0} & 7,306 \\
        \bottomrule
    \end{tabular}
    \vspace{-8pt}
\end{table}

\textbf{Effects of encoder fine-tuning.}
Figure~\ref{fig:frozen_vs_finetuned} shows that node-encoder fine-tuning increases the separation between gold-overlapping and non-gold regions across text, tables, and video.
Measured directly on test gold items, fine-tuning raises the share of correct parent--child refinement decisions from 28.3--43.0\% to 34.3--71.9\% across benchmarks (Appendix~\ref{app:decision_accuracy}).
Table~\ref{tab:frozen_vs_finetuned} reports the downstream effects without routing.
Images and tables answering aggregate questions are always shown whole and left out of the share of pruned items.
Fine-tuning raises the share of pruned items from 30.4--39.6\% to 85.2--95.6\%, reducing reader-input evidence tokens by 6.8--19.1\% while changing F1/accuracy by $+0.3$ to $+0.9$ points, all of whose 95\% CIs include zero ($[-0.2, +2.1]$, $[-1.0, +1.5]$, $[-0.8, +2.5]$, $[-1.2, +2.2]$, and $[-2.0, +3.0]$ on NQ, HotpotQA, OTT-QA, MMQA, and LVBench).

\begin{figure}[H]
    \vspace{-6pt}
    \centering
    \begin{minipage}[c]{0.57\textwidth}
    \centering
    \includegraphics[width=\linewidth]{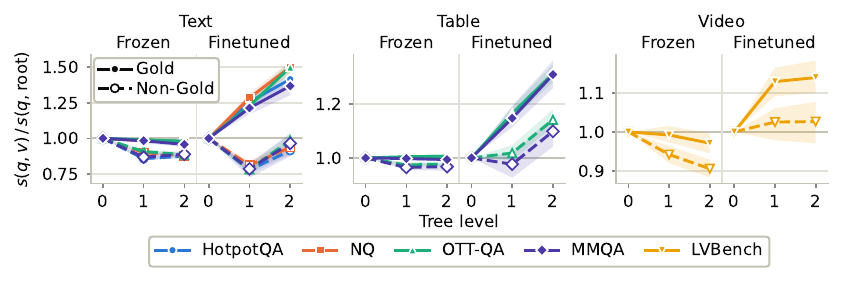}
    \caption{Median root-relative node scores, $s(q,v)/s(q,\text{root})$, at each tree level for gold-overlapping nodes (Gold, solid) and other nodes (Non-Gold, dashed), using the frozen encoder $f_0$ and fine-tuned node encoder $f_\theta$. Shaded bands show 95\% bootstrap confidence intervals for the medians.}
    \label{fig:frozen_vs_finetuned}
    \end{minipage}
    \hfill
    \begin{minipage}[c]{0.40\textwidth}
    \centering
    \captionsetup{skip=4pt}
    \captionof{table}{\canopy{} without routing, using the frozen ($f_0$) or fine-tuned ($f_\theta$) node encoder. Pruned is the share of items from which refinement removes some content.}
    \label{tab:frozen_vs_finetuned}
    \scriptsize
    \setlength{\tabcolsep}{3pt}
    \renewcommand{\arraystretch}{1.15}
    \resizebox{\linewidth}{!}{%
    \begin{tabular}{@{}lrr>{\columncolor{gray!12}}r>{\columncolor{gray!12}}rrr@{}}
        \toprule
        \multirow{2}{*}{Benchmark} & \multicolumn{2}{c}{Pruned (\%)} & \multicolumn{2}{c}{\#Tok} & \multicolumn{2}{c}{F1 / Acc.} \\
        \cmidrule(lr){2-3} \cmidrule(lr){4-5} \cmidrule(lr){6-7}
        & $f_0$ & $f_\theta$ & $f_0$ & $f_\theta$ & $f_0$ & $f_\theta$ \\
        \midrule
        NQ & 38.6 & 95.0 & 1,664 & 1,512 & 53.5 & \textbf{54.4} \\
        HotpotQA & 39.6 & 95.6 & 1,983 & 1,816 & 61.8 & \textbf{62.1} \\
        OTT-QA & 38.7 & 85.2 & 3,001 & 2,796 & 32.5 & \textbf{33.4} \\
        MMQA & 38.1 & 90.1 & 2,304 & 2,123 & 50.7 & \textbf{51.1} \\
        LVBench & 30.4 & 90.9 & 4,918 & 3,978 & 34.4 & \textbf{35.0} \\
        \bottomrule
    \end{tabular}}
    \end{minipage}
    \vspace{-4pt}
\end{figure}

%% file: 5_conclusion.tex
\section{Conclusion}
\label{sec:conclusion}

We studied post-retrieval evidence compression for heterogeneous multimodal RAG, focusing on adaptively determining the extent and granularity of evidence retained within retrieved items.
We introduced \canopy{}, which combines a node encoder fine-tuned on gold evidence with a shared parent--child embedding similarity rule to select multiple regions at different granularities without LLM calls for node-level pruning.
We couple this compression with critic-guided additional retrieval to acquire missing information while limiting the evidence volume accumulated across retrieval rounds. Across five QA benchmarks over a heterogeneous corpus of approximately 33M items, \canopy{} achieves higher average answer accuracy than the evaluated retrieval baselines.
Component ablations indicate that additional retrieval drives the main accuracy gains on multi-hop QA, while compression reduces reader-input evidence volume within the same iterative retrieval pipeline.
In the unrouted Qwen3-VL-8B-Instruct setting, compression reduces reader-input evidence tokens by 14.2--27.7\% relative to the same pipeline without compression, with comparable answer accuracy.
These results demonstrate that a shared adaptive compression procedure can reduce reader input across heterogeneous retrieval results.

%% file: 99_appendix.tex
\section{Dataset and Corpus Details}
\label{app:corpus}

Table~\ref{tab:dataset_stats} summarizes the five in-domain benchmarks of Section~\ref{sec:experiments} and the three out-of-domain benchmarks of Appendix~\ref{app:ood}: the modalities in which their gold evidence resides, the number of questions we use, the retrieval collection each contributes to its pool, and the metric.
All queries are text-only.
For every in-domain benchmark, we start from its development split and keep the questions whose gold evidence can be located in the corpus: 3,563 of NQ's 7,830, 7,404 of HotpotQA's 7,405, all 2,214 of OTT-QA, 2,423 of MMQA's 2,441, and 662 of LVBench's 777.
NQ shrinks the most because we exclude questions without a short answer and require the gold evidence to map onto a DPR passage.
We then randomly split each set with a fixed seed into disjoint test and train portions, stratified by reasoning type where the benchmark provides one (bridge and comparison questions for HotpotQA, and the 14 original question types for MMQA).
The test portion has 1,000 questions (500 for LVBench); up to 1,000 of the remaining questions (162 for LVBench) form the train portion, which is used only to fine-tune the node encoder (Appendix~\ref{app:implementation_details}).
The out-of-domain benchmarks follow the same conventions but contribute no train portion, since nothing is fine-tuned or tuned on them.

\subsection{In-Domain Benchmarks}

\paragraph{Natural Questions (NQ).}
NQ~\citep{kwiatkowski2019natural} consists of real user queries issued to Google Search, with answers annotated on supporting Wikipedia pages.
It serves as our single-hop text QA benchmark.
Of the 7,830 development questions, we keep the 3,563 that have a short answer and whose gold evidence maps onto a DPR passage, split them randomly, and retrieve over the 21.0M Wikipedia passages of DPR~\citep{karpukhin2020dense}, each of at most 100 words.
Answers are short spans, evaluated with EM and F1.

\paragraph{HotpotQA.}
HotpotQA~\citep{yang2018hotpotqa} is a Wikipedia-based benchmark whose questions require reasoning over two articles, with the supporting sentences annotated as supporting facts.
We use the fullwiki setting, so retrieval runs over the 5.2M Wikipedia abstracts rather than a per-question candidate set.
We use 7,404 of the 7,405 development questions, split with stratification over bridge and comparison questions.
Supporting facts provide sentence-level gold spans; answers are evaluated with EM and F1.

\paragraph{OTT-QA.}
OTT-QA~\citep{chen2021ottqa} is an open-domain benchmark over tables and text: each question is grounded in a Wikipedia table and typically requires a passage linked from one of its cells.
Unlike in its closed-domain predecessor, no table or passage is given with the question, so both must be retrieved; the corpus contributes the 419K tables and the 6.0M passages linked from them.
We use all 2,214 development questions, split randomly.
Answers are evaluated with EM and F1.

\paragraph{MultimodalQA (MMQA).}
MMQA~\citep{talmor2021multimodalqa} contains questions over text, tables, and images, each annotated with the modalities its reasoning requires.
It contributes 218K passages, 10K tables, and 57K images to the corpus, and the annotated answer instances (table cells, passage spans, and image identifiers) provide gold spans.
We use 2,423 of the 2,441 development questions, split with stratification over the benchmark's 14 question types.
Answers are evaluated with EM and F1.

\paragraph{LVBench.}
LVBench~\citep{wang2025lvbench} is a long-video understanding benchmark with multiple-choice questions on YouTube videos averaging over an hour.
We use the adaptation released by UniversalRAG~\citep{yeo2026universalrag}, which rephrases the original video-interleaved questions into text-only queries and annotates each with a time interval on its video; of the released queries, 662 refer to the 77 videos that remain available, and we randomly use 500 of them for testing and the remaining 162 for fine-tuning.
The corpus contributes the 1,529 pre-cut segments of these videos, each at most five minutes long.
Answers are evaluated with accuracy.

\subsection{Out-of-Domain Benchmarks}

\paragraph{2WikiMultiHopQA.}
2WikiMultiHopQA~\citep{ho2020twowiki} asks two-hop questions over Wikipedia (compositional, comparison, bridge-comparison, and inference types), with the supporting sentences annotated.
We use the packaging of UniversalRAG~\citep{yeo2026universalrag}: 2,000 questions sampled from the dev split, whose candidate paragraphs form the collection.
An item is one paragraph (11,678 distinct), titled by its article, and the supporting facts give sentence-level gold spans.
Answers are evaluated with EM and F1.

\paragraph{TAT-QA.}
TAT-QA~\citep{zhu2021tatqa} asks about a page of a financial report that holds one table and its paragraphs, with the supporting paragraphs annotated; we keep its span and multi-span questions, whose answers are located, and leave out the arithmetic ones.
Its questions presuppose the page (``What was the net income in 2019?''), and the pages carry no identifier, so we use T$^2$-RAGBench~\citep{strich2025t2ragbench}, whose test pages are the TAT-QA test pages: its context-independent rewrites of the questions (419 of the 924 test span questions, matched to the TAT-QA annotations by page and answer, so the gold cells and sentences are kept) and its company name and report year, which title every page (``Spirent Communications Plc 2019 annual report'').
The collection is the 227 TAT-QA pages of these questions, each one table with its paragraphs (227 tables, 1,034 paragraphs), and the 2,446 T$^2$-RAGBench train and dev pages, which are whole report pages that hold several tables each (5,386 tables, 25,206 paragraphs); every table and every paragraph is one item, 5,613 and 26,240 in all.
Answers are evaluated with EM and F1.

\paragraph{LongVideoBench.}
LongVideoBench~\citep{wu2024longvideobench} asks multiple-choice questions about YouTube videos of up to an hour, each question referring to a moment of its video that the annotators mark by frame index.
We use the validation split and keep the 815 questions of the ten visual categories, leaving out the seven that refer to or ask about subtitles, which the reader does not receive; 733 of them carry a usable frame index.
Their 455 videos are cut into 1,489 segments of at most three minutes, one item each, as for LVBench, and the referred frame, widened to $\pm$15 seconds on its segment's clock, is the gold span.
Answers are evaluated with accuracy.

\begin{table}[H]
    \centering
    \caption{Benchmark and corpus summary. Used lists the questions kept out of the split they come from. The corpus columns give the items each benchmark contributes to its pool by modality: the unified corpus of 32.9M items for the in-domain benchmarks (Section~\ref{sec:experiments}), and the separate 45,020-item pool of the out-of-domain benchmarks (Appendix~\ref{app:ood}), which contribute no train portion.}
    \label{tab:dataset_stats}
    \footnotesize
    \setlength{\tabcolsep}{4pt}
    \renewcommand{\arraystretch}{1.15}
    \resizebox{\textwidth}{!}{%
    \begin{tabular}{@{}llrrrrrrrc@{}}
        \toprule
        \multirow{2}{*}{Benchmark} & \multirow{2}{*}{Evidence modality} & \multirow{2}{*}{Used / split} & \multirow{2}{*}{\#Test} & \multirow{2}{*}{\#Train} & \multicolumn{4}{c}{Corpus items} & \multirow{2}{*}{Metric} \\
        \cmidrule(lr){6-9}
        & & & & & Text & Table & Image & Video & \\
        \midrule
        \rowcolor{gray!20} \multicolumn{10}{c}{\emph{In-domain}} \\
        NQ & Text & 3,563 / 7,830 dev & 1,000 & 1,000 & 21.0M & -- & -- & -- & EM / F1 \\
        HotpotQA & Text (multi-hop) & 7,404 / 7,405 dev & 1,000 & 1,000 & 5.2M & -- & -- & -- & EM / F1 \\
        OTT-QA & Table + Text & 2,214 / 2,214 dev & 1,000 & 1,000 & 6.0M & 419K & -- & -- & EM / F1 \\
        MMQA & Text + Table + Image & 2,423 / 2,441 dev & 1,000 & 1,000 & 218K & 10K & 57K & -- & EM / F1 \\
        LVBench & Video & 662 / 777 dev & 500 & 162 & -- & -- & -- & 1.5K & Acc. \\
        \cmidrule(lr){1-10}
        Total & & & 4,500 & 4,162 & 32.4M & 429K & 57K & 1.5K & \\
        \midrule
        \rowcolor{gray!20} \multicolumn{10}{c}{\emph{Out-of-domain}} \\
        2WikiMultiHopQA & Text (multi-hop) & 2,000 / 12,576 dev & 2,000 & -- & 11,678 & -- & -- & -- & EM / F1 \\
        TAT-QA & Table + Text & 419 / 924 test span & 419 & -- & 26,240 & 5,613 & -- & -- & EM / F1 \\
        LongVideoBench & Video & 733 / 1,337 val & 733 & -- & -- & -- & -- & 1,489 & Acc. \\
        \cmidrule(lr){1-10}
        Total & & & 3,152 & -- & 37,918 & 5,613 & -- & 1,489 & \\
        \bottomrule
    \end{tabular}%
    }
\end{table}

\subsection{Corpus Construction and Gold Spans}

\paragraph{In-domain corpus.}
The unified corpus combines the retrieval collections released with the five in-domain benchmarks (Table~\ref{tab:dataset_stats}).
Text items are the Wikipedia passages of DPR~\citep{karpukhin2020dense} used for NQ, the Wikipedia abstracts of the HotpotQA fullwiki setting, the passages linked from OTT-QA tables, and the MMQA passages; table items are the OTT-QA and MMQA Wikipedia tables; image items are the MMQA images; video items are the 1,529 pre-cut segments of the 77 LVBench videos in the packaging released by UniversalRAG~\citep{yeo2026universalrag}, each at most five minutes long.
Every item is embedded once with Qwen3-VL-Embedding-2B: text and table items from their title followed by their content, with tables flattened one row per line, images from their pixels, and video segments from frames sampled at one frame per 10 seconds (at most 32).
The corpus holds 32.9M items in a single flat FAISS index with fp16 scalar quantization and exact inner-product search; text accounts for 32.4M of them, tables for 429K, images for 57K, and video segments for 1.5K.

\paragraph{Out-of-domain corpus.}
The three out-of-domain collections are merged into a pool of their own, kept separate from the unified corpus, which every out-of-domain query searches in full (Appendix~\ref{app:ood}).
It holds 45,020 items: 37,918 paragraphs, 5,613 tables, and 1,489 video segments.
Items are formed and embedded exactly as above, with the same encoder and the same flat index, so the only difference from the in-domain setting is the content of the pool.

\paragraph{Gold span annotation.}
Gold spans are the fragments of a gold item that support the answer; we use the benchmark's own annotations where they exist and derive them from the answer otherwise, as described above for the out-of-domain benchmarks and below for the in-domain ones.
HotpotQA provides supporting facts as sentence indices, which we use directly.
For LVBench, we use the \texttt{time\_reference} annotations released with UniversalRAG's adaptation~\citep{yeo2026universalrag}, which give a time interval in seconds on the full video, and map each interval onto the pre-cut segments and their 30-second clips by half-open interval overlap.
The gold video that AKS and \canopy{} receive in Section~\ref{sec:setup} is the pre-cut segment that contains this interval.
MMQA provides the positions of answer instances, namely table cells, character spans in passages, and image identifiers; cells and spans map to the rows and sentences containing them, images are whole items, and cells linking to another supporting document are added as bridge cells.
OTT-QA traces the answer to table cells and linked passages; the cells map to rows, and within a passage the gold span is the sentences containing the answer string.
NQ annotates answers on the original Wikipedia page rather than on DPR passages, so we select the passage under the same title that overlaps the long answer most and take the sentences containing the short answer as the gold span, or the whole passage when no sentence matches.
Supporting items without a located span are treated as whole-item gold and yield no fine-tuning instances.

\section{Implementation Details}
\label{app:implementation_details}

This section completes the setup summarized in Section~\ref{sec:experiments}: node-encoder fine-tuning, the routed variant and context caps of \canopy{}, baseline reproduction, latency measurement, and how the metrics are counted.

The node encoder $f_\theta$ is initialized from the shared retrieval encoder Qwen3-VL-Embedding-2B and fine-tuned with the objective of Section~\ref{sec:encoder-fine-tuning}, using only the train portion of each benchmark, which is disjoint from the test questions.
We use up to 1,000 training questions per benchmark (162 for LVBench), 4,162 in total, whose gold spans (Appendix~\ref{app:corpus}) yield $|\mathcal D|=18{,}575$ training instances $(q,v)$.
These instances are pooled across the five benchmarks to train a single LoRA adapter, which is shared across benchmarks.
LoRA uses rank 8, $\alpha=16$, and dropout 0.05, and is applied to the attention and MLP projections of the language backbone; the vision tower is frozen.
We set $\lambda=20$ in Equation~\ref{eq:node-ranking-loss} and train with AdamW at a learning rate of $5\times10^{-5}$ and a batch size of 16 instances for 2 epochs (2,322 steps).

At inference, \canopy{}$^\dagger$ routes every query, including each follow-up query, with the modality router in Appendix~\ref{app:prompts} and, as in UniversalRAG, retrieves $k$ items from each routed corpus.
Reader inputs are capped at 16k tokens on NQ, HotpotQA, OTT-QA, and MMQA; on LVBench, whose retrieved videos rarely fit in a short context, the cap is 64k for Qwen3-VL-8B-Instruct and 40k for InternVL3.5-8B.
Evidence beyond the cap is truncated latest-retrieved first, for both the critic and the reader.

Follow-up queries often retrieve items that are already in the accumulated evidence.
Such items are not compressed again: they remain in the evidence with the regions selected when they were first retrieved.
Of the items returned by follow-up retrieval, 63.3\% for \canopy{} and 62.8\% for \canopy{}$^\dagger$ were already in the evidence (51.2--76.6\% and 51.9--74.2\% across benchmarks).

Baselines are run with publicly released implementations under the same retrieval encoder, corpus, and reader.
For RECOMP and S2G-RAG on OTT-QA, we treat each table row as a sentence.
S2G-RAG originally uses a trained sufficiency and gap judge; in our reproduction, this judge is replaced by the same untrained LLM that serves as the reader, so that every method uses one LLM for all of its calls.
In IRCoT, at each of at most three iterations, the reader generates one reasoning sentence, which becomes the next retrieval query, and the chain stops once it states ``So the answer is:''.
A chain that reaches this statement is answered by the text that follows it, and one that has not reached it after three iterations is answered, as in the original IRCoT, by the reader over all documents it has gathered.

Lat.\ is the wall-clock time of query encoding, routing, projection, and every LLM call over a batch, including prompt construction and queueing on the client, divided by the number of questions; it is an amortized time under batched execution, not the latency of an individual question.
Vector search and node encoding are excluded.
All methods, including UniversalRAG and IRCoT, are timed under the same serving conditions: questions are processed in batches of 200 with 32 requests in flight against four vLLM servers, one per NVIDIA RTX A6000 (48\,GB), with prefix caching disabled.

\#Tok counts the evidence tokens in the reader's answer prompt with the reader's tokenizer and image preprocessing, for every method, and LLM Calls includes the final answer call.
Where differences between methods are small, we report 95\% paired bootstrap confidence intervals of the difference over test questions (10,000 resamples).
Averages, ratios, and pooled rates are computed from unrounded values and rounded only for display.

\section{Algorithm of \canopy{}}
\label{app:algorithm}

Algorithm~\ref{alg:canopy} gives the pseudocode for the hierarchical refinement of a single retrieved item described in Section~\ref{sec:canonical-projection}.
Here, $q$ is the query used in the current retrieval round, as specified in Section~\ref{sec:critic}.
The prompts for the LLM components used alongside this procedure, namely the critic (Section~\ref{sec:critic}), the aggregate classifier for retrieved tables, and the modality router of \canopy{}$^\dagger$, are listed in Appendix~\ref{app:prompts}.

\begin{algorithm}[H]
\caption{\canopy{}: Hierarchical Evidence Compression}
\label{alg:canopy}
\small
\begin{algorithmic}[1]
\Require Retrieval query $q$, retrieved item $x_i$, branching factor $B$, similarity score $s(q,v)$
\Ensure Selected node set $\mathcal S_i$, the roots of the evidence forest for $x_i$

\State Construct hierarchy $\mathcal T_i$ for $x_i$ as in Section~\ref{sec:canonical-projection}
\State $\mathcal S_i\gets\{\operatorname{root}(\mathcal T_i)\}$
\While{$\mathcal S_i$ contains an unprocessed non-leaf node}
    \State Choose such a node $v$ and mark it as processed
    \If{some child $u$ of $v$ satisfies $s(q,u)\geq s(q,v)$}
        \State $\mathcal S_i\gets\mathcal S_i\setminus\{v\}$
        \ForAll{children $u$ of $v$ satisfying $s(q,u)\geq s(q,v)$}
            \State $\mathcal S_i\gets\mathcal S_i\cup\{u\}$
        \EndFor
        \State Discard the remaining child subtrees
    \Else
        \State Retain $v$ in $\mathcal S_i$ without further refinement
    \EndIf
\EndWhile
\State \Return $\mathcal S_i$
\end{algorithmic}
\end{algorithm}

\section{Compression Methods at Other Evidence Budgets}
\label{app:compression_budget}

\begin{table}[H]
    \centering
    \caption{AKS at different frame budgets on LVBench, where both methods receive the gold video instead of retrieved items. $M$: frames selected per video, whose default is $M=16$.}
    \label{tab:compression_budget_video}
    \scriptsize
    \setlength{\tabcolsep}{5pt}
    \renewcommand{\arraystretch}{1.15}
    \begin{tabular}{@{}llr>{\columncolor{gray!12}}r@{}}
        \toprule
        \multirow{2}{*}{Method} & \multirow{2}{*}{Budget} & \multicolumn{2}{c}{LVBench} \\
        \cmidrule(lr){3-4}
        & & Acc. & \#Tok \\
        \midrule
        \multirow{3}{*}{AKS} & $M=8$ & 48.0 & 1,010 \\
        & $M=16$ & 52.2 & 2,024 \\
        & $M=32$ & 53.6 & 4,054 \\
        \rowcolor{ours} \canopy{} & --- & \textbf{56.4} & 2,782 \\
        \bottomrule
    \end{tabular}
\end{table}

\begin{table}[H]
    \centering
    \caption{Compression methods at larger evidence budgets, using each benchmark's own
    retrieval corpus. $N$: retained units for RECOMP-extractive, whose default is $N=1$;
    $\rho$: target compression rate for LongLLMLingua, whose default is $0.55$.}
    \label{tab:compression_budget}
    \scriptsize
    \setlength{\tabcolsep}{2.5pt}
    \renewcommand{\arraystretch}{1.15}
    \begin{tabular}{@{}llrr>{\columncolor{gray!12}}rrr>{\columncolor{gray!12}}rrr>{\columncolor{gray!12}}r@{}}
        \toprule
        \multirow{2}{*}{Method} & \multirow{2}{*}{Budget}
        & \multicolumn{3}{c}{NQ} & \multicolumn{3}{c}{HotpotQA} & \multicolumn{3}{c}{OTT-QA} \\
        \cmidrule(lr){3-5} \cmidrule(lr){6-8} \cmidrule(lr){9-11}
        & & EM & F1 & \#Tok & EM & F1 & \#Tok & EM & F1 & \#Tok \\
        \midrule
        \multirow{5}{*}{RECOMP-extractive} & $N=1$ & 27.0 & 38.8 & 37 & 32.6 & 43.0 & 41 & 10.3 & 14.8 & 43 \\
        & $N=5$ & 31.3 & 44.2 & 184 & 35.9 & 48.6 & 200 & 10.6 & 15.2 & 223 \\
        & $N=10$ & 32.2 & 45.7 & 359 & 37.2 & 50.0 & 392 & 10.7 & 15.8 & 452 \\
        & $N=20$ & 30.7 & 44.3 & 701 & 35.2 & 49.1 & 749 & 12.4 & 17.5 & 905 \\
        & $N=50$ & 29.6 & 44.2 & 1,573 & 34.9 & 49.0 & 1,133 & 12.3 & 17.8 & 2,187 \\
        \midrule
        \multirow{4}{*}{LongLLMLingua} & $\rho=0.55$ & 34.1 & 49.7 & 805 & 40.3 & 50.9 & 572 & 12.5 & 17.0 & 1,651 \\
        & $\rho=0.70$ & 35.3 & 50.8 & 1,038 & 39.8 & 50.7 & 731 & 12.5 & 16.7 & 2,123 \\
        & $\rho=0.90$ & 36.3 & 52.2 & 1,331 & 40.0 & 51.2 & 945 & 13.5 & 18.2 & 2,774 \\
        & $\rho=0.95$ & 36.2 & 52.2 & 1,383 & 39.7 & 51.2 & 985 & 13.3 & 17.9 & 2,936 \\
        \rowcolor{ours} \canopy{} w/o Iterative & --- & 37.4 & 52.2 & 1,382 & 41.8 & 52.3 & 1,003 & 13.6 & 18.3 & 2,196 \\
        \rowcolor{ours} \canopy{} & --- & \textbf{38.7} & \textbf{53.6} & 1,555 & \textbf{52.0} & \textbf{64.3} & 1,352 & \textbf{29.0} & \textbf{34.7} & 3,150 \\
        \bottomrule
    \end{tabular}
\end{table}

\section{Additional Component Ablations with Modality Routing}
\label{app:routed_ablation}

Table~\ref{tab:component_ablation_routed} extends the component ablation in Table~\ref{tab:component_ablation} to modality-routed retrieval.
Compression reduces evidence tokens on all five benchmarks, but its answer-quality effects remain dataset-dependent.
The largest accuracy loss occurs on LVBench: tokens decrease from 40,365 to 26,907 (33.3\%), while accuracy falls from 45.2 to 43.4.
Removing iterative retrieval lowers HotpotQA and OTT-QA EM from 47.6 to 38.0 and from 20.1 to 11.5, respectively.

\begin{table}[H]
    \centering
    \caption{Effect of \canopy{}'s components with modality routing.
    $^\dagger$Queries are routed to modality-specific corpora before retrieval.
    \canopy{}$^\dagger$ is reproduced from Table~\ref{tab:overall_results}; the other rows are measured separately.
    Lat.\ is the processing time per question in seconds under batched execution (Appendix~\ref{app:implementation_details}).}
    \label{tab:component_ablation_routed}
    \footnotesize
    \setlength{\tabcolsep}{2pt}
    \renewcommand{\arraystretch}{1.15}
    \resizebox{\textwidth}{!}{%
    \begin{tabular}{@{}lrr>{\columncolor{gray!12}}r>{\columncolor{gray!12}}rrr>{\columncolor{gray!12}}r>{\columncolor{gray!12}}rrr>{\columncolor{gray!12}}r>{\columncolor{gray!12}}rrr>{\columncolor{gray!12}}r>{\columncolor{gray!12}}rc>{\columncolor{gray!12}}r>{\columncolor{gray!12}}r@{}}
        \toprule
        \multirow{2}{*}{Method}
        & \multicolumn{4}{c}{NQ} & \multicolumn{4}{c}{HotpotQA} & \multicolumn{4}{c}{OTT-QA} & \multicolumn{4}{c}{MMQA} & \multicolumn{3}{c}{LVBench} \\
        \cmidrule(lr){2-5} \cmidrule(lr){6-9} \cmidrule(lr){10-13} \cmidrule(lr){14-17} \cmidrule(lr){18-20}
        & EM & F1 & \#Tok & Lat. & EM & F1 & \#Tok & Lat. & EM & F1 & \#Tok & Lat. & EM & F1 & \#Tok & Lat. & Acc. & \#Tok & Lat. \\
        \midrule
        \rowcolor{ours} \canopy{}$^\dagger$ & 38.1 & \textbf{52.7} & 2,715 & 0.64 & \textbf{47.6} & \textbf{60.1} & 2,967 & 0.59 & 20.1 & 24.9 & 4,929 & 0.95 & \textbf{42.1} & \textbf{47.4} & 4,251 & 1.23 & 43.4 & 26,907 & 6.54 \\
        w/o Projection & \textbf{38.3} & 52.4 & 2,872 & 0.66 & 46.9 & 59.1 & 3,494 & 0.63 & \textbf{21.3} & \textbf{26.4} & 5,563 & 0.99 & 41.4 & 47.0 & 4,649 & 1.25 & \textbf{45.2} & 40,365 & 13.18 \\
        w/o Iterative & 36.9 & 51.7 & 2,323 & 0.22 & 38.0 & 48.7 & 2,156 & 0.16 & 11.5 & 15.4 & 3,368 & 0.22 & 38.3 & 43.5 & 3,118 & 0.34 & 44.0 & 21,364 & 1.75 \\
        w/o Projection, Iterative & 37.8 & 52.1 & 2,548 & 0.22 & 37.7 & 48.3 & 2,523 & 0.17 & 11.4 & 15.7 & 3,862 & 0.23 & 38.3 & 43.8 & 3,400 & 0.34 & 44.4 & 33,267 & 4.42 \\
        \bottomrule
    \end{tabular}%
    }
\end{table}

\section{Sensitivity to Branching Factor and Retrieval Size}
\label{app:parameter_sensitivity}

\paragraph{Branching factor.}
Table~\ref{tab:branching_factor} varies $B\in\{2,3,4\}$ at $k=10$.
Within each routing configuration, EM varies by at most 1.0 point on NQ, HotpotQA, and MMQA, while the largest changes occur in OTT-QA EM without routing (1.9 points) and LVBench accuracy with routing (4.0 points).
Without routing, increasing $B$ from 2 to 4 lowers OTT-QA EM from 27.7 to 25.8 and raises LVBench accuracy from 35.0 to 35.4 (95\% CI of the difference $[-2.2, +3.0]$).
With routing, $B=3$ gives the highest LVBench accuracy (47.4).
A larger branching factor therefore does not consistently improve accuracy or reduce evidence volume; we use $B=2$ as the default.

\paragraph{Retrieval size.}
Table~\ref{tab:topk} varies $k\in\{5,10,20\}$ for \canopy{} at $B=2$ and compares it with the retrieval baselines.
Without routing, increasing $k$ from 5 to 20 increases \canopy{}'s evidence tokens 3.6--3.9$\times$, while EM changes by at most 2.8 points and LVBench accuracy stops improving after $k=10$ (95\% CI of the difference between $k=20$ and $k=10$ $[-3.2, +3.2]$).
Average retrieval rounds remain between 2.0 and 2.1.
The routed setting also shows substantially higher token usage with limited or non-monotonic quality changes as $k$ increases.
We use $k=10$ as the default.

\paragraph{Pooled results.}
Figure~\ref{fig:topk_pooled} summarizes the Qwen3-VL-8B-Instruct results in Table~\ref{tab:topk}.
Both axes are means over the five benchmarks.
Answer quality uses EM on the first four benchmarks and accuracy on LVBench; evidence tokens are accumulated over retrieval rounds.

\paragraph{Comparison with retrieval baselines.}
Without routing, \canopy{} with $k=5$ exceeds Vanilla RAG with $k=20$ on HotpotQA, OTT-QA, and MMQA while using only 31--35\% as many evidence tokens.
This comparison evaluates the complete \canopy{} pipeline, including additional retrieval, rather than isolating compression.
On NQ, \canopy{} with $k=10$ approaches Vanilla RAG with $k=20$ (39.7 versus 39.8 EM; 95\% CI of the difference $[-1.6, +1.4]$) at about half the tokens.
LVBench is an exception: unrouted Vanilla RAG with $k=20$ reaches 36.2 accuracy, exceeding unrouted \canopy{} at every tested $k$.

\begin{table}[H]
    \centering
    \caption{Effect of the branching factor $B$ on \canopy{} at $k=10$, with and without modality routing.}
    \label{tab:branching_factor}
    \scriptsize
    \setlength{\tabcolsep}{2pt}
    \renewcommand{\arraystretch}{1.1}
    \begin{tabular}{@{}ccrr>{\columncolor{gray!12}}rrr>{\columncolor{gray!12}}rrr>{\columncolor{gray!12}}rrr>{\columncolor{gray!12}}rc>{\columncolor{gray!12}}r@{}}
        \toprule
        \multirow{2}{*}{Routing} & \multirow{2}{*}{$B$}
        & \multicolumn{3}{c}{NQ} & \multicolumn{3}{c}{HotpotQA} & \multicolumn{3}{c}{OTT-QA} & \multicolumn{3}{c}{MMQA} & \multicolumn{2}{c}{LVBench} \\
        \cmidrule(lr){3-5} \cmidrule(lr){6-8} \cmidrule(lr){9-11} \cmidrule(lr){12-14} \cmidrule(lr){15-16}
        & & EM & F1 & \#Tok & EM & F1 & \#Tok & EM & F1 & \#Tok & EM & F1 & \#Tok & Acc. & \#Tok \\
        \midrule
        \multirow{3}{*}{w/o} & 2 & \textbf{39.7} & \textbf{54.4} & 1,512 & 49.6 & 62.1 & 1,816 & \textbf{27.7} & \textbf{33.4} & 2,796 & 45.1 & 51.1 & 2,123 & 35.0 & 3,978 \\
         & 3 & 39.0 & 53.5 & 1,560 & \textbf{49.8} & \textbf{62.3} & 1,862 & 26.9 & 32.9 & 2,765 & \textbf{45.2} & \textbf{51.2} & 2,130 & 34.6 & 4,117 \\
         & 4 & 39.2 & 53.9 & 1,538 & 49.6 & \textbf{62.3} & 1,827 & 25.8 & 31.6 & 2,734 & 44.6 & 50.8 & 2,101 & \textbf{35.4} & 4,312 \\
        \midrule
        \multirow{3}{*}{w/} & 2 & 38.1 & 52.7 & 2,715 & 47.6 & \textbf{60.1} & 2,967 & \textbf{20.1} & \textbf{24.9} & 4,929 & \textbf{42.1} & \textbf{47.4} & 4,251 & 43.4 & 26,907 \\
         & 3 & 38.2 & 52.5 & 2,714 & \textbf{47.7} & 59.8 & 3,021 & 19.8 & 24.5 & 4,828 & 41.5 & 47.1 & 4,265 & \textbf{47.4} & 27,257 \\
         & 4 & \textbf{38.4} & \textbf{53.0} & 2,745 & 47.6 & 59.6 & 2,960 & 19.8 & 24.7 & 4,695 & 41.1 & 46.5 & 4,212 & 46.0 & 27,617 \\
        \bottomrule
    \end{tabular}
\end{table}

\begin{table}[H]
    \centering
    \caption{Effect of retrieval size $k$ on Vanilla RAG, UniversalRAG, IRCoT, and \canopy{}. We fix $B=2$ for \canopy{}. Vanilla RAG and UniversalRAG use single-round retrieval, while \canopy{} is shown with and without routing and retrieves again when evidence is insufficient. For UniversalRAG, $k$ is the number of items retrieved per routed corpus.
    Bold marks the highest EM, F1, and accuracy at each $k$.}
    \label{tab:topk}
    \scriptsize
    \setlength{\tabcolsep}{2pt}
    \renewcommand{\arraystretch}{1.15}
    \resizebox{\textwidth}{!}{%
    \begin{tabular}{@{}ccrr>{\columncolor{gray!12}}rrr>{\columncolor{gray!12}}rrr>{\columncolor{gray!12}}rrr>{\columncolor{gray!12}}rc>{\columncolor{gray!12}}rr@{}}
        \toprule
        \multirow{2}{*}{Method} & \multirow{2}{*}{$k$}
        & \multicolumn{3}{c}{NQ} & \multicolumn{3}{c}{HotpotQA} & \multicolumn{3}{c}{OTT-QA} & \multicolumn{3}{c}{MMQA} & \multicolumn{2}{c}{LVBench}
        & \multirow{2}{*}{LLM Calls} \\
        \cmidrule(lr){3-5} \cmidrule(lr){6-8} \cmidrule(lr){9-11} \cmidrule(lr){12-14} \cmidrule(lr){15-16}
        & & EM & F1 & \#Tok & EM & F1 & \#Tok & EM & F1 & \#Tok & EM & F1 & \#Tok & Acc. & \#Tok & \\
        \midrule
        \multirow{3}{*}{Vanilla RAG} & 5 & 36.8 & 50.9 & 795 & 37.3 & 47.5 & 748 & 13.5 & 17.6 & 1,147 & 37.6 & 42.7 & 977 & 33.6 & 2,338 & 1.0 \\
        & 10 & 38.2 & 52.6 & 1,565 & 38.9 & 49.7 & 1,520 & 12.9 & 17.5 & 2,185 & 40.1 & 45.9 & 1,863 & 34.4 & 4,236 & 1.0 \\
        & 20 & 39.8 & 54.7 & 3,083 & 41.3 & 51.7 & 3,070 & 14.4 & 19.4 & 4,133 & 40.4 & 46.1 & 3,609 & 36.2 & 7,681 & 1.0 \\
        \midrule
        \multirow{3}{*}{UniversalRAG} & 5 & 32.9 & 47.8 & 1,734 & 35.2 & 45.4 & 2,117 & 10.6 & 14.4 & 3,049 & 34.5 & 39.3 & 2,353 & 40.0 & 17,258 & 2.0 \\
        & 10 & 35.0 & 50.4 & 2,804 & 36.5 & 46.7 & 3,980 & 10.5 & 14.4 & 5,546 & 36.9 & 42.0 & 4,131 & 41.8 & 33,853 & 2.0 \\
        & 20 & 34.8 & 50.1 & 4,600 & 37.2 & 47.9 & 7,238 & 11.0 & 15.6 & 9,408 & 37.5 & 43.1 & 7,218 & 42.4 & 54,216 & 2.0 \\
        \midrule
        \multirow{3}{*}{IRCoT} & 5 & 33.7 & 50.3 & 1,669 & 42.7 & 58.0 & 1,681 & 22.6 & 29.1 & 2,419 & 37.0 & 45.1 & 1,971 & 17.0 & 4,671 & 3.7 \\
        & 10 & 35.0 & 51.3 & 3,179 & 47.4 & 61.8 & 3,363 & 24.3 & 30.8 & 4,707 & 38.9 & 46.9 & 3,817 & 17.8 & 8,518 & 3.6 \\
        & 20 & 36.4 & 52.6 & 5,993 & 49.9 & \textbf{63.7} & 6,879 & 25.6 & 31.7 & 9,040 & 39.9 & 47.4 & 7,436 & 16.4 & 14,343 & 3.5 \\
        \midrule
        \rowcolor{ours} & \multicolumn{16}{@{}l}{\textit{w/o Routing}} \\
        \rowcolor{ours} & 5 & \textbf{37.5} & \textbf{51.9} & 756 & \textbf{49.5} & \textbf{61.2} & 946 & \textbf{26.3} & \textbf{31.6} & 1,458 & \textbf{44.0} & \textbf{49.2} & 1,106 & 33.0 & 2,051 & 2.9 \\
        \rowcolor{ours} & 10 & \textbf{39.7} & \textbf{54.4} & 1,512 & \textbf{49.6} & \textbf{62.1} & 1,816 & \textbf{27.7} & \textbf{33.4} & 2,796 & \textbf{45.1} & \textbf{51.1} & 2,123 & 35.0 & 3,978 & 3.0 \\
        \rowcolor{ours} & 20 & \textbf{40.3} & \textbf{55.8} & 2,982 & \textbf{50.7} & 63.0 & 3,555 & \textbf{27.6} & \textbf{33.2} & 5,362 & \textbf{44.0} & \textbf{50.3} & 4,064 & 35.0 & 7,306 & 3.0 \\
        \rowcolor{ours} & \multicolumn{16}{@{}l}{\textit{w/ Routing}} \\
        \rowcolor{ours} & 5 & 36.5 & 50.8 & 1,685 & 49.0 & 60.7 & 1,696 & 19.6 & 24.1 & 2,659 & 40.8 & 45.9 & 2,564 & \textbf{43.2} & 13,421 & 4.9 \\
        \rowcolor{ours} & 10 & 38.1 & 52.7 & 2,715 & 47.6 & 60.1 & 2,967 & 20.1 & 24.9 & 4,929 & 42.1 & 47.4 & 4,251 & \textbf{43.4} & 26,907 & 4.9 \\
        \rowcolor{ours} \multirow{-8}{*}{\canopy{}} & 20 & 40.1 & 55.4 & 4,345 & 49.2 & 61.0 & 5,341 & 20.2 & 24.8 & 8,607 & 41.9 & 47.8 & 7,199 & \textbf{44.2} & 50,072 & 4.7 \\
        \bottomrule
    \end{tabular}%
    }
\end{table}

\section{Flat versus Hierarchical Selection with the Fine-Tuned Encoder}
\label{app:projection_strategies}

\begin{figure}[H]
    \centering
    \includegraphics[width=0.5\linewidth]{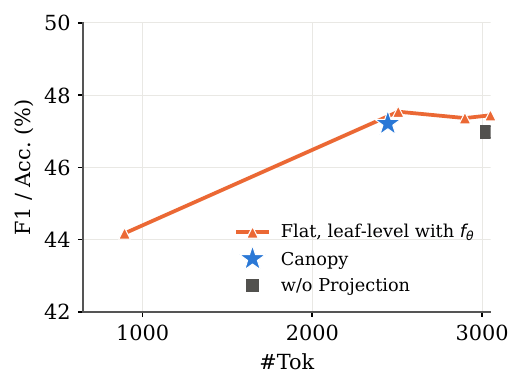}
    \caption{Pooled answer quality versus evidence tokens for leaf-level flat selection with $f_\theta$ at each $m$, \canopy{}, and \textit{w/o Projection}, without modality routing.}
    \label{fig:topm_unit}
\end{figure}

The comparison in Section~\ref{sec:ablation} scores flat selection with the retrieval encoder $f_0$, so it does not separate the contribution of the fine-tuned node encoder $f_\theta$ from that of hierarchical refinement.
Here we give flat selection the same fine-tuned encoder as \canopy{}, so that the two differ only in the selection rule, and ask what the traversal itself adds.
Leaf-level flat selection scores the leaves of each retrieved item with $f_\theta$ and retains the $m$ highest-scoring ones.
The rest of the pipeline is unchanged, and we sweep $m\in\{1,5,10,20\}$ without modality routing, using Qwen3-VL-8B-Instruct.
Figure~\ref{fig:topm_unit} plots pooled answer quality, the mean of F1 on the four text-based benchmarks and accuracy on LVBench, against pooled evidence tokens, the mean over the five benchmarks.

With the fine-tuned encoder, flat selection traces a quality--token curve that rises steeply for small $m$ and then plateaus, while tokens keep growing up to the volume of \textit{w/o Projection}.
\canopy{} lies at the knee of this curve, with token volume and quality comparable to the smallest $m$ on the plateau.
This comparison therefore does not establish a quality--token advantage over the well-performing settings of the flat selector.
What differs is how the retained extent is set: flat selection needs $m$ chosen by sweeping with the reader, and the chosen $m$ applies the same maximum retained-unit count to every item, whereas \canopy{} decides the extent per item through parent-relative branch selection and stopping, without an externally specified count.
The two are not equivalent in general: retaining a parent keeps a lower-scoring leaf while a higher-scoring leaf in another branch is discarded, a selection that no prefix of the leaf ranking reproduces.
As an illustration with made-up scores, let a root with score 0.50 have children $A$ (0.80) with leaves $a_1$ (0.79) and $a_2$ (0.10), and $B$ (0.70) with leaves $b_1$ (0.90) and $b_2$ (0.20).
Refinement descends into both children, keeps $A$ whole because neither leaf reaches 0.80, and replaces $B$ by $b_1$, retaining $\{a_1, a_2, b_1\}$; the leaf ranking $b_1 > a_1 > b_2 > a_2$ admits no $m$ that retains $a_2$ without $b_2$.
Such retained parents are common in practice: they account for 25--35\% of the refinement decisions in Appendix~\ref{app:refinement_anatomy}.
Together with the encoder ablation in Table~\ref{tab:frozen_vs_finetuned} and the decision accuracy in Figure~\ref{fig:decision_accuracy}, these results support parent-relative refinement as a learned procedure shared across heterogeneous items, rather than as a better operating point.

\section{Accuracy of Parent--Child Refinement Decisions}
\label{app:decision_accuracy}

The root-relative scores in Figure~\ref{fig:frozen_vs_finetuned} show that fine-tuning separates gold from non-gold regions, but they do not directly measure whether the refinement rule makes the right choice at a node.
We therefore evaluate the parent--child comparisons themselves on the gold items of the test questions.
The decision instances are defined exactly as in training (Section~\ref{sec:encoder-fine-tuning}): every node $v \in \mathcal{V}_i$, that is, every gold-overlapping internal node without a fully gold proper ancestor.
Each instance is scored independently of retrieval and of the traversal path, so the measurement isolates the encoder from the rest of the pipeline.
A decision at $v$ is counted as correct only when every preference in $\mathcal{P}(v)$ holds (Section~\ref{sec:encoder-fine-tuning}): gold-overlapping children must score at least as high as $v$ and the remaining children strictly below it, or, when $v$ is fully gold, all children must score strictly below $v$.
Images are excluded because they are never split.

\begin{figure}[H]
    \centering
    \includegraphics[width=0.85\linewidth]{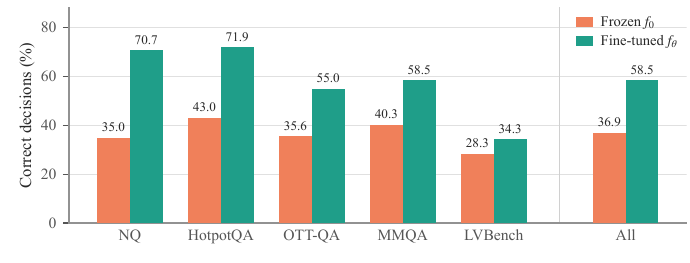}
    \caption{Accuracy of parent--child refinement decisions with the frozen ($f_0$) and fine-tuned ($f_\theta$) node encoder on the gold items of the test questions.}
    \label{fig:decision_accuracy}
\end{figure}

Figure~\ref{fig:decision_accuracy} reports the share of correct decisions.
With the frozen encoder $f_0$, between 28.3\% and 43.0\% of decisions are correct, so parent-relative similarity alone is an unreliable refinement signal.
Fine-tuning raises the share on every benchmark, to 58.5\% overall (from 36.9\%), with the largest gains on NQ (35.0 to 70.7) and HotpotQA (43.0 to 71.9), and smaller gains on OTT-QA (35.6 to 55.0) and MMQA (40.3 to 58.5).
LVBench improves the least, from 28.3\% to 34.3\%, consistent with the weaker gold/non-gold separation for video in Figure~\ref{fig:frozen_vs_finetuned}.
Because a decision is correct only when all of its preferences hold, this criterion is strict; a single misordered child at a node with several children counts as an error even when \canopy{} would still retain the gold region.

We also measure gold-fragment recall in the full pipeline: for each question, over its gold items with a located gold span among the shown items that refinement may split, in the evidence of the round the reader answers on, the share of their gold fragments in $G_i$ that remain after refinement, averaged over the questions with such items.
Images and tables answering aggregate questions are always shown whole and left out.
Gold-fragment recall is 79.4\% for \canopy{} and 77.7\% for \canopy{}$^\dagger$, against 100\% when the items are shown whole; Table~\ref{tab:gold_retention} breaks it down by benchmark.

\begin{table}[H]
    \centering
    \caption{Gold-fragment recall after refinement for \canopy{} and \canopy{}$^\dagger$, over the gold items with a located gold span among the shown items that refinement may split.
    Recall is computed per question over its gold items and averaged over questions.}
    \label{tab:gold_retention}
    \scriptsize
    \setlength{\tabcolsep}{6pt}
    \renewcommand{\arraystretch}{1.15}
    \begin{tabular}{@{}lrr@{}}
        \toprule
        Benchmark & \canopy{} (\%) & \canopy{}$^\dagger$ (\%) \\
        \midrule
        NQ & 76.1 & 76.2 \\
        HotpotQA & 81.9 & 80.9 \\
        OTT-QA & 79.1 & 78.4 \\
        MMQA & 84.1 & 85.2 \\
        LVBench & 60.7 & 74.3 \\
        \midrule
        All & 79.4 & 77.7 \\
        \bottomrule
    \end{tabular}
\end{table}

\section{How Refinement Proceeds}
\label{app:refinement_anatomy}

\begin{figure}[H]
    \centering
    \includegraphics[width=\linewidth]{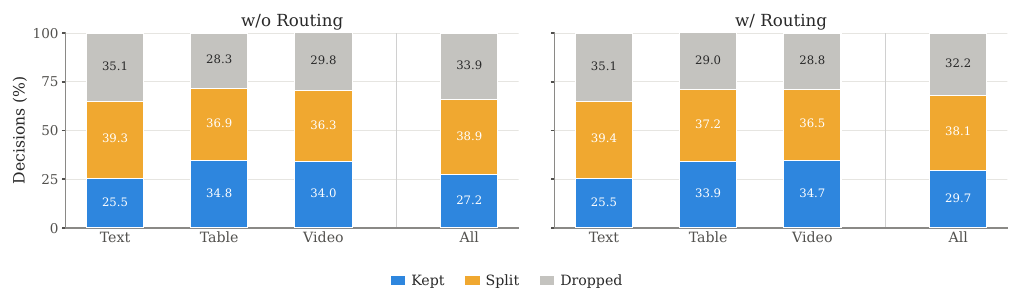}
    \caption{What refinement decides at each node: kept, split, or dropped, by modality,
    for \canopy{} and \canopy{}$^\dagger$, over the shown items that refinement may split.
    Items are those in the evidence of the round the reader answers on, counted once per question; ``All'' pools the decisions across modalities.}
    \label{fig:node_fate}
\end{figure}

This section describes what refinement decides over the shown items that refinement may split, for \canopy{} and \canopy{}$^\dagger$; images and tables answering aggregate questions are always shown whole and left out.
Figure~\ref{fig:node_fate} looks at refinement one node at a time, by modality.
Each decision is one of the three outcomes of the refinement rule (Section~\ref{sec:canonical-projection}): the node is \emph{kept} when no child qualifies and it is retained, \emph{split} when it is replaced by its qualifying children, and \emph{dropped} when it is a non-qualifying child discarded at its parent's split; we read the outcome off the shown fragments, of which all, some, or none remain.
Text is refined more aggressively than tables and video: 25.5\% of the text decisions keep a node and 35.1\% drop one, against roughly a third kept and 28--30\% dropped for tables and video, and the shares barely move with routing.
A text item is a sequence of sentences of which a question typically needs one or two, whereas the rows of a table and the segments of a video more often carry the answer together, so the parent--child comparison finds a child that stands out more often in text.

\section{Refinement on Non-Gold Items}
\label{app:nongold_refinement}

\begin{figure}[H]
    \centering
    \includegraphics[width=\linewidth]{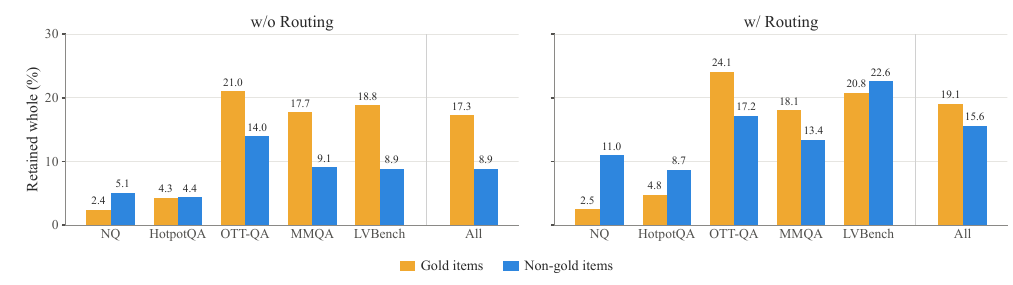}
    \caption{Share of items that refinement keeps whole, that is, stops at the root,
    split by whether an item is gold, for \canopy{} and \canopy{}$^\dagger$, over the shown items that refinement may split; ``All'' pools the items across the benchmarks.}
    \label{fig:retained_whole}
\end{figure}

The node encoder is fine-tuned on preference pairs built from gold-overlapping nodes (Section~\ref{sec:encoder-fine-tuning}), so we check how refinement behaves on gold and non-gold items at test time.
We measure the share of items that refinement keeps whole, that is, stops at the root, separately for gold and non-gold items, for both \canopy{} and \canopy{}$^\dagger$.
Figure~\ref{fig:retained_whole} reports these rates: \canopy{} keeps 8.9\% of non-gold and 17.3\% of gold items whole, and \canopy{}$^\dagger$ keeps 15.6\% and 19.1\%.
Overall, non-gold items are refined below the root more often than gold items, although this varies across benchmarks, so refinement is not confined to items that support the answer.

\section{Out-of-Domain Evaluation}
\label{app:ood}

The node encoder $f_\theta$ and every setting of \canopy{} are fixed on the five in-domain benchmarks of Table~\ref{tab:dataset_stats}.
To check that they carry over, we repeat the Table~\ref{tab:overall_results} comparison on three benchmarks the pipeline never saw, one for text, one for text with tables, and one for video (Appendix~\ref{app:corpus}), with the corpus built the same way: their collections merged into one pool that every query searches.
Everything else follows Table~\ref{tab:overall_results}: $k=10$, three rounds, the same reader, $f_\theta$ from Appendix~\ref{app:implementation_details}, and IRCoT with ten documents and up to three iterations.

\begin{table}[H]
    \centering
    \caption{Out-of-domain results over one pool of the three benchmarks' collections, in the configuration of Table~\ref{tab:overall_results}.
    Avg.\ is the mean of the two EM values and the accuracy; LLM Calls is the mean per question.
    \#Tok is the evidence the reader actually receives.
    $^\dagger$Queries are routed to modality-specific parts of the pool before retrieval, following UniversalRAG.}
    \label{tab:ood_unified}
    \footnotesize
    \setlength{\tabcolsep}{4pt}
    \renewcommand{\arraystretch}{1.15}
    \resizebox{0.70\textwidth}{!}{%
    \begin{tabular}{@{}lrr>{\columncolor{gray!12}}rrr>{\columncolor{gray!12}}rr>{\columncolor{gray!12}}rcr@{}}
        \toprule
        \multirow{2}{*}{Method}
        & \multicolumn{3}{c}{2WikiMultiHopQA} & \multicolumn{3}{c}{TAT-QA} & \multicolumn{2}{c}{LongVideoBench} & \multirow{2}{*}{Avg.} & \multirow{2}{*}{LLM Calls} \\
        \cmidrule(lr){2-4} \cmidrule(lr){5-7} \cmidrule(lr){8-9}
        & EM & F1 & \#Tok & EM & F1 & \#Tok & Acc. & \#Tok & & \\
        \midrule
        Vanilla RAG & 44.1 & 50.2 & 1,388 & \textbf{71.4} & \textbf{72.4} & 1,273 & \textbf{49.5} & 57,736 & 55.0 & 1.0 \\
        UniversalRAG & 42.8 & 48.9 & 1,673 & 58.5 & 59.5 & 1,819 & 47.1 & 41,554 & 49.4 & 2.0 \\
        IRCoT & \textbf{53.2} & \textbf{63.8} & 3,328 & 60.1 & 64.7 & 2,788 & 35.1 & 61,746 & 49.5 & 3.6 \\
        \rowcolor{ours} \canopy{} & 51.0 & 60.2 & 1,587 & 70.9 & 72.1 & 1,220 & 48.2 & 57,195 & \textbf{56.7} & 2.7 \\
        \rowcolor{ours} \canopy{}$^\dagger$ & 50.2 & 59.4 & 1,944 & 59.2 & 60.4 & 1,794 & 47.2 & 50,320 & 52.2 & 4.8 \\
        \bottomrule
    \end{tabular}%
    }
\end{table}

\canopy{} has the best average of the five methods, 1.7 above Vanilla RAG, with the node encoder and every setting carried over unchanged from the five in-domain benchmarks of Table~\ref{tab:dataset_stats}.
Benchmark by benchmark, Table~\ref{tab:ood_unified} repeats the pattern of Table~\ref{tab:overall_results}: gains from additional retrieval on the multi-hop benchmark, and parity with single-round retrieval at equal or lower token cost on the other two.
On 2WikiMultiHopQA, where the initial top-10 holds both hops for only 47\% of the questions, \canopy{} improves over Vanilla RAG by 10.0 F1, almost entirely on the compositional questions (47.0 vs.\ 25.7 F1), whose second hop the critic retrieves in a later round.
As in the ablation of Section~\ref{sec:ablation}, the gain coincides with additional retrieval while the evidence volume stays close to single-round retrieval: IRCoT, the other iterative method, is 3.6 F1 ahead of \canopy{} but spends 2.1$\times$ the evidence tokens and 3.6 calls per question, so \canopy{} reaches most of the multi-hop gain on 48\% of IRCoT's tokens and stays within 15\% of single-round Vanilla RAG's.
On TAT-QA and LongVideoBench, the evidence is concentrated in one page or one segment, so additional retrieval has less to add and the projection is judged mainly on what it removes.
Unlike on LVBench over the unified corpus, unrouted retrieval on LongVideoBench returns mostly video segments, as the token counts show.
On both benchmarks, \canopy{} stays within 1.3 points of single-round Vanilla RAG (72.1 vs.\ 72.4 F1 and 48.2 vs.\ 49.5 accuracy) while passing 4\% fewer evidence tokens on TAT-QA (1,220 vs.\ 1,273) and 0.9\% fewer on LongVideoBench (57,195 vs.\ 57,736), whereas IRCoT, which accumulates whole items over its iterations, loses 7.7 F1 on TAT-QA and falls to 35.1 on LongVideoBench, where a quarter of its answers to the multiple-choice questions come as free text rather than as an option letter.
TAT-QA's tables are financial report pages, unlike the Wikipedia tables the node encoder was fine-tuned on, so the row-level projection carries over to a table distribution it has not seen: it cuts the pages it is given at a 0.3 F1 difference from passing them whole.
LongVideoBench, whose videos were not seen in fine-tuning, shows the same: \canopy{} stays within 1.3 points of Vanilla RAG at 0.9\% fewer evidence tokens.
Modality routing costs 12--13 F1 on TAT-QA for UniversalRAG and \canopy{}$^\dagger$ alike: the router sends 84\% of the questions to the tables alone, where the answers that sit in the paragraphs are lost (57.6 vs.\ 71.6 F1 on those questions); on LongVideoBench it sends 30\% of the questions to the text pool, which costs 8 points on them; and on 2WikiMultiHopQA it adds calls and evidence tokens without improving answer quality.

\section{Case Study}
\label{app:case_study}

Tables~\ref{tab:case_study} and~\ref{tab:case_study_app} compare \canopy{} with Vanilla RAG on two multi-hop questions.
Vanilla RAG passes every retrieved item to the reader in full within a single round, whereas \canopy{} retains the regions of each item selected by parent-relative refinement and, when its critic judges the evidence insufficient, retrieves again with a query for the missing information.

\textbf{Acquiring missing evidence across modalities.}
Table~\ref{tab:case_study} illustrates how compression and additional retrieval contribute at different stages.
The initial results identify David Douillet's sport but do not state when it was invented.
Within the UNESCO Champion for Sport table, \canopy{} retains three of thirteen rows, including the row linking Douillet to judo.
This preserves the intermediate fact needed for the next reasoning step, without reducing the table to a single answer-bearing cell.
The critic then identifies the missing invention date and issues the targeted query ``When was judo invented?''
From the subsequently retrieved Judo passage, \canopy{} retains two of eight sentences, including the statement that judo was created in 1882.
The case illustrates how compressed evidence can support a follow-up query that retrieves information absent from the initial results.

\textbf{Adapting the retained context to each item.}
Table~\ref{tab:case_study_app} highlights that useful evidence need not be compressed to the same extent across items.
\canopy{} retains the initial Center for Architecture passage in full, preserving the link between the institution and Greenwich Village.
The critic recognizes that this establishes the relevant neighborhood but leaves the transformation of its Dutch name unresolved.
A follow-up query retrieves the Greenwich Village article, from which \canopy{} retains two of eight sentences, including the explanation that ``Groenwijck'' was Anglicized to ``Greenwich.''
Thus, the evidence used to answer the question combines an intact passage with a smaller region from another item.
This example illustrates the flexibility of retaining broader context where needed while refining other items, with the critic determining which additional fact remains to be retrieved.

\begin{table}[H]
    \centering
    \caption{Case study comparing \canopy{} with Vanilla RAG: acquiring missing evidence across modalities.}
    \label{tab:case_study}
    \footnotesize
    \renewcommand{\arraystretch}{1.2}
    \begin{tabularx}{\textwidth}{>{\raggedright\arraybackslash}p{0.17\textwidth}@{\hspace{4mm}}>{\raggedright\arraybackslash}X}
        \toprule
        \textbf{Question} & {\small What year was the sport invented for the sport that David Douillet, current UNESCO Champion for Sport, plays?}\\
        \midrule
        \textbf{Ground Truth} & {\small 1882}\\
        \midrule
        \textbf{Vanilla RAG} & \textbf{Retrieved:} ten items, all shown in full: the UNESCO Champion for Sport table (13 rows, including ``David Douillet $|$ France $|$ 2002 $|$ \textcolor{blue!70!black}{judo}'') and nine passages on David Douillet and the UNESCO Champion for Sport, none of which states when judo was invented.\\
        & \textbf{Response:} 1964\;\textcolor{red!70!black}{\ding{55}}\\
        \midrule
        \textbf{\canopy{} (Ours)} & \textit{Round 1}\\
        & \textbf{Retrieved:} the same ten items (1,424 $\rightarrow$ 1,013 tokens). Of the UNESCO Champion for Sport table, \canopy{} retains 3 of 13 rows (350 $\rightarrow$ 126 tokens):\\
        & \begin{minipage}[t]{\linewidth}
            \centering
            \scriptsize
            \begin{tabular}{@{}llll@{}}
                \toprule
                \textbf{Name} & \textbf{Country} & \textbf{Since} & \textbf{Sport}\\
                \midrule
                \textcolor{blue!70!black}{David Douillet} & France & 2002 & \textcolor{blue!70!black}{judo}\\
                Vyacheslav Fetisov & Russia & 26 May 2004 & ice hockey\\
                Hakuho (Davaajargal Munkhbat) & Mongolia & 27 July 2006 & sumo wrestling\\
                \bottomrule
            \end{tabular}
        \end{minipage}\\
        & \textbf{Critic:} ``The evidence confirms David Douillet is a judoka and a UNESCO Champion for Sport, but it does not state when judo was invented.''\\
        & \textbf{Query:} ``When was judo invented?''\\
        \cmidrule(l){2-2}
        & \textit{Round 2}\\
        & \textbf{Retrieved:} ten new items (1,728 $\rightarrow$ 1,428 tokens), including the Judo article, of which \canopy{} retains 2 of 8 sentences (282 $\rightarrow$ 110 tokens): ``The sport was originally created in \textcolor{blue!70!black}{1882} by Jigoro Kano \ldots{} as a physical, mental, and moral pedagogy in Japan. Judo's most prominent feature is its competitive element, \ldots{} force an opponent to submit with a joint lock or a choke.''\\
        & \textbf{Critic:} ``The evidence states David Douillet is a judoka and that judo was created in 1882 by Jigoro Kano.''\\
        & \textbf{Response:} 1882\;\textcolor{green!50!black}{\cmark}\\
        \bottomrule
    \end{tabularx}
\end{table}

\begin{table}[H]
    \centering
    \caption{Case study comparing \canopy{} with Vanilla RAG: adapting the retained context to each item.}
    \label{tab:case_study_app}
    \footnotesize
    \renewcommand{\arraystretch}{1.2}
    \begin{tabularx}{\textwidth}{>{\raggedright\arraybackslash}p{0.17\textwidth}@{\hspace{4mm}}>{\raggedright\arraybackslash}X}
        \toprule
        \textbf{Question} & {\small What was done to the Dutch name of the NYC neighborhood that contains the Center for Architecture to form its present name?}\\
        \midrule
        \textbf{Ground Truth} & {\small Anglicized}\\
        \midrule
        \textbf{Vanilla RAG} & \textbf{Retrieved:} ten items, all shown in full, including ``The Center for Architecture is located in the neighborhood of \textcolor{blue!70!black}{Greenwich Village} \ldots'' and ``The Archipelbuurt (literally `Archipelago Neighbourhood') is a neighbourhood in the Centrum district of The Hague, Netherlands.'' None of them gives the Dutch name of Greenwich Village.\\
        & \textbf{Response:} It was not changed; the neighborhood is called Greenwich Village.\;\textcolor{red!70!black}{\ding{55}}\\
        \midrule
        \textbf{\canopy{} (Ours)} & \textit{Round 1}\\
        & \textbf{Retrieved:} the same ten items (1,540 $\rightarrow$ 1,233 tokens). \canopy{} retains the Center for Architecture passage whole: ``The Center for Architecture is located in the neighborhood of \textcolor{blue!70!black}{Greenwich Village} at 536 LaGuardia Place, \ldots''\\
        & \textbf{Critic:} ``The evidence identifies the Center for Architecture as being in Greenwich Village, Manhattan, but does not mention any Dutch name or linguistic transformation for this neighborhood.''\\
        & \textbf{Query:} ``What is the Dutch name of Greenwich Village and how was it changed to `Greenwich Village'?''\\
        \cmidrule(l){2-2}
        & \textit{Round 2}\\
        & \textbf{Retrieved:} ten new items (1,701 $\rightarrow$ 1,370 tokens), including the Greenwich Village article, of which \canopy{} retains 2 of 8 sentences (409 $\rightarrow$ 78 tokens): ``Greenwich Village also contains several subsections, including the West Village west of Seventh Avenue and the Meatpacking District in the northwest corner of Greenwich Village. The neighborhood's name comes from \textcolor{blue!70!black}{Groenwijck}, one of the Dutch names for the village (meaning Green District), which was \textcolor{blue!70!black}{Anglicized} to Greenwich.''\\
        & \textbf{Critic:} ``The evidence states the Dutch name `Groenwijck' was Anglicized to `Greenwich,' and the Center for Architecture is located in Greenwich Village, which is part of NYC.''\\
        & \textbf{Response:} Anglicized\;\textcolor{green!50!black}{\cmark}\\
        \bottomrule
    \end{tabularx}
\end{table}

\section{Prompts}
\label{app:prompts}

The critic (Section~\ref{sec:critic}) judges whether the accumulated evidence suffices and, if not, issues a follow-up query using the prompt below.

\begin{promptbox}{Critic Prompt}
You are given numbered evidence and a question.\newline
Work in order and reply with exactly five lines and nothing else:\newline
RATIONALE: <one sentence: what the evidence establishes toward the question, and what it does not>\newline
ANSWER: <your best answer from the evidence alone, or NONE if it does not support one>\newline
SUFFICIENT: <YES if that answer is supported by the evidence, otherwise NO>\newline
MISSING: <if NO, the single fact the evidence still does not give; if YES, NONE>\newline
QUERY: <if NO, a search query for that missing fact; if YES, NONE>\newline
If the question lists lettered options, ANSWER is that option's letter alone.\newline
Say YES whenever the answer can be read off the evidence, even if it is not exhaustive.\newline
Any gap listed under "Already searched" was searched for and not found. Do not name it again: either the answer is already present, in which case say YES, or a different fact is missing.\newline
Do not explain.
\end{promptbox}

Once a table appears among a question's retrieved items, the question alone is classified once with the prompt below; for an aggregate question, \canopy{} passes retrieved tables whole instead of pruning them.

\begin{promptbox}{Aggregate Classifier Prompt}
Classify the following query as aggregate or lookup, based on how much of its table is needed to answer it. The table is not shown; judge from the query alone. Consider:\newline
- aggregate: the answer needs rows the query does not name: counting or listing rows that meet a condition, a superlative or rank over a table column (earliest year, lowest attendance), comparing against other rows, or a condition checked over every row.\newline
- lookup: the query names the one row it needs, by a name, title, year or description, possibly asking a further fact about it. A "how many" or superlative about that one entity is still lookup.\par\medskip
A wrong aggregate only costs extra tokens; a wrong lookup can drop the row that holds the answer. If the query ranks, counts or collects rows before the answering row is known, choose aggregate.\par\medskip
Examples:\newline
1. "What was the earliest year that actor Steve Burton was nominated for his work on General Hospital by the Soap Opera Digest Award Association?" -> aggregate\newline
2. "What role did Dick Cusack play in the film titled Crazy People?" -> lookup\newline
3. "Which 2019 films that Apoorva Arora acted in are in the Hindi language?" -> aggregate\newline
4. "How many people live in the hometown of the 2012-13 LEB Oro team that was founded in 1966 ?" -> lookup\newline
5. "What is the population of the city in which the Armenian Premier League stadium with the smallest capacity is located ?" -> aggregate\newline
6. "The 2018 Sydney International seed player from the fourth-largest country in the Americas is how tall in centimetres ?" -> lookup\par\medskip
Classify the following query: \{query\}\newline
Provide only the label: aggregate or lookup.
\end{promptbox}

The router used by \canopy{}$^\dagger$ and the w/ Routing variants is adapted from UniversalRAG~\citep{yeo2026universalrag}. Since \canopy{} determines granularity after retrieval, the original paragraph/document and clip/video categories are merged into text and video, respectively.

\begin{promptbox}{Modality Router Prompt}
Classify the following query into one or more categories from: [text, table, image, video], based on whether it requires retrieval-augmented generation (RAG) and the most appropriate modality. Consider:\newline
- text: The query requires retrieving factual descriptions, straightforward explanations, or concise summaries, whether from a single source or by combining information from multiple documents.\newline
- table: The query requires information that is best represented in a tabular format, often involving comparisons or structured data.\newline
- image: The query focuses on visual aspects like appearances, structures, or spatial relationships.\newline
- video: The query targets a short, specific moment or event within a video, or requires understanding dynamic events, motion, or sequences over time.\par\medskip
For cross-modality queries that require multiple types of information, combine categories with '+' (e.g., text+image).\par\medskip
Examples:\newline
1. "What is the birth date of Alan Turing?" -> text\newline
2. "Which academic discipline do computer scientist Alan Turing and mathematician John von Neumann have in common?" -> text\newline
3. "Among the recipients of the Turing Award, who had the earliest birth year?" -> table\newline
4. "Describe the appearance of a blue whale." -> image\newline
5. "Describe the moment Messi scored his goal in the 2022 World Cup final." -> video\newline
6. "Explain how Messi scored his goal in the 2022 World Cup final." -> video\newline
7. "Who played a key role in the development of the iPhone?" -> text\newline
8. "Which Harvard University graduate played a key role in the development of the iPhone?" -> text\newline
9. "What is the cheapest iPhone model available in 2023?" -> table\newline
10. "Describe the structure of the Eiffel Tower." -> image\newline
11. "Describe the moment Darth Vader reveals he is Luke's father in Star Wars." -> video\newline
12. "Analyze the sequence of events leading to the fall of the Empire in Star Wars." -> video\newline
13. "Describe the visual appearance and habitat of the blue whale." -> text+image\newline
14. "Compare the architectural features shown in Gothic and Renaissance cathedrals." -> image+table\newline
15. "Describe the moment of the moon landing and explain the mission details." -> text+video\par\medskip
Classify the following query: \{query\}\newline
Provide only the category or categories combined with '+'.
\end{promptbox}